\documentclass[superscriptaddress, aps,pre, twocolumn, floatfix]{revtex4-2}

\usepackage{graphicx}
\usepackage{xcolor}
\usepackage{physics}
\usepackage[utf8]{inputenc}
\usepackage[T1]{fontenc}
\usepackage{float}
\usepackage[english]{babel}
\usepackage{epsfig}
\usepackage{epstopdf}
\usepackage{amsmath}
\usepackage{amssymb}

\usepackage{enumitem}
\usepackage{mathptmx}
\usepackage{braket}

\usepackage[colorlinks=true, linkcolor=blue,citecolor=red, urlcolor=blue, hypertexnames=true]{hyperref}

\newcommand{\eref}[1]{Eq.~\eqref{#1}}
\newcommand{\Eref}[1]{Equation~\eqref{#1}}
\newcommand{\aref}[1]{Appendix~\ref{#1}}
\newcommand{\sref}[1]{Sec.~\ref{#1}}
\newcommand{\Sref}[1]{Section~\ref{#1}}
\newcommand{\fref}[1]{Fig.~\ref{#1}}
\newcommand{\Fref}[1]{Figure~\ref{#1}}

\begin{document}
\title{Sluggish quantum mechanics of noninteracting fermions with spatially varying effective mass}
\date{\today}

\author{Giuseppe Del Vecchio Del Vecchio}
\email{giuseppedelvecchiodelvecchio@gmail.com}
\affiliation{Laboratoire de Physique de l’\'{E}cole Normale Sup\'{e}rieure, ENS, Universit\'{e} PSL,
CNRS, Sorbonne Universit\'{e}, Universit\'{e} Paris Cit\'{e}, 75005 Paris, France}

\author{Manas Kulkarni}
\email{manas.kulkarni@icts.res.in}
\affiliation{International Centre for Theoretical Sciences, Tata Institute of Fundamental Research,
Bangalore 560089, India}

\author{Satya N. Majumdar}
\email{satyanarayan.majumdar@cnrs.fr}
\affiliation{LPTMS, CNRS, Universit\'e Paris-Sud, Universit\'e Paris-Saclay, 91405 Orsay, France}

\author{Sanjib Sabhapandit}
\email{sanjib@rri.res.in}
\affiliation{Raman Research Institute, Bangalore 560080, India}

\begin{abstract}
We analyze a class of one-dimensional quantum systems characterized by a position-dependent kinetic term arising as the continuum limit of an inhomogeneous tight-binding model with spatially varying hopping amplitudes. In this limit, the Schr\"odinger equation takes the so-called BenDaniel–Duke form with an effective mass, scaling as 
$m_\mathrm{eff}(x) = m_\mathrm{eff}\,|x|^{\alpha}$
with $\alpha > 0$, leading to a framework we term ``sluggish quantum mechanics'', where particle motion is progressively suppressed at larger distances. For the case with no external confinement, we obtain the eigenfunctions and the quantum propagator exactly. In the presence of a confining potential $V_\mathrm{ext}(x)=\frac{1}{2} m_\mathrm{eff}(x)\,\omega^2 x^2 =\frac{1}{2}m_\mathrm{eff}\,\omega^2 |x|^{\alpha+2}$, we show that the model remains exactly solvable, and obtain the spectral properties and the propagator analytically. We then investigate the problem of $N$ noninteracting spinless fermions in the trap, determining the many-body ground-state wavefunction and the joint probability density function of the positions of the $N$ fermions. 
We show that the many-body quantum probability density in the ground state forms a 
determinantal point process whose correlation kernel can be computed for any $N$, giving 
access to the average density as well as higher order correlation functions for any finite 
$N$. Moreover, we analyze the scaling form of this kernel in the large $N$ limit in the 
bulk, near the edge, and close to the origin. Our results show that the scaled average 
density profile for large $N$ has a finite support symmetric with respect to the origin, but 
has a non-monotonic shape with a vanishing minimum at the origin for any $\alpha>0$.
One of the key findings of our work is that the scaled kernel near the origin $x=0$ for $\alpha>0$ is neither the Bessel nor the Airy kernel (that are standard for trapped fermions), but is new, and is given by a sum of two Bessel kernels with different indices. Our results thus provide a framework relevant to engineered optical lattices with position-dependent tunneling.
\end{abstract}

\maketitle

\section{Introduction}

Advances in ultracold atom experiments have made it possible to realize various one-dimensional quantum lattice models on optical lattices~\cite{BDZ2008,Giorgini2008, BDN2012, LSA2012, GAN2014, GB2017,Bloch2018}. These quantum lattice models are generically described by tight-binding type Hamiltonians that are characterized by an onsite potential and hopping amplitude across the bonds. A system with a uniform onsite potential,   hopping amplitude, and a lattice spacing $\delta$, is usually constructed by using two counter-propagating laser beams, whose interference pattern forms a cosine potential $V(x)=V_0\cos^2(\pi x/\delta)$, where $V_0$ is the depth of the potential~\cite{Jaksch1998,MO2006}. Interestingly, space-dependent (non-uniform) onsite potentials $\epsilon_i$ have also been experimentally realized, which form test-beds for understanding various exotic phenomena such as localization and phase transition~\cite{Aspect2008,Roati_2008,Lahini2009, Schreiber2015}. Moreover, using precise programmable techniques, one can engineer both space-dependent onsite energies and bond-dependent tunneling~\cite{Gauthier2016, Qiu2020, Wei2024, Tabares2025}.  

When the external potential $V(x)$ consists of a sequence of deep potential wells, whose minima are separated by some characteristic length scale $\delta$, it is very useful and justified to approximate the Schr\"odinger equation in terms of an effective localized basis. In this localized basis, the quantum particle tunnels from one minimum to another, effectively, making a quantum lattice model with a lattice spacing $\delta$. In the tight-binding approximation, tunneling only between nearest neighbors is considered. By suitably considering a space-dependent hopping element, taking the limit of lattice spacing tending to zero and hopping strength to infinity, one can arrive at a Schrodinger equation for the coarse-grained lattice description, with a space-dependent effective mass $m_\text{eff}(x)$ as (see \sref{s:derivation})
\begin{equation}
   i\hbar \frac{\partial \psi(x,t) }{\partial t}   =- \frac{\partial}{\partial x} \left(\frac{\hbar^2}{2 m_\text{eff}(x)} \frac{\partial \psi (x,t)}{\partial x}  \right)  .
  \label{eq:SE-slug0i}  
\end{equation}

The operator $-\partial_x(\hbar^2/[2m_\text{eff}(x)]\partial_x)$ in
\eref{eq:SE-slug0i} generalizes the standard kinetic-energy operator
$-[\hbar^2/(2m)]\partial_x^2$ to the case of a spatially varying effective
mass. 
Interestingly, the Schr\"odinger equation in \eqref{eq:SE-slug0i},
with a spatially dependent effective mass, corresponds to the
BenDaniel--Duke form~\cite{BenDaniel-Duke1966}. This operator arises
naturally in the study of electron tunneling across semimetal and
semiconductor heterojunctions. Alternative Hermitian (self-adjoint)
kinetic-energy operators with position-dependent effective mass have
also been proposed and analyzed in related contexts~\cite{Gora1969,Bastard1975,Zhu1983,Roos1983,harrison2016quantum,osti_5095386}.
In most studies of semiconductor heterostructures, however, the effective mass is assumed to be piecewise constant, i.e., uniform
within each layer and discontinuous only at interfaces. 
In this paper, we consider the spatially varying effective mass of the form
$m_{\mathrm{eff}}(x)= m_\mathrm{eff}\, |x|^{\alpha}$, where $\alpha\ge 0$ and $m_\mathrm{eff}\,$ on the right-hand side is a constant. This may be envisioned as the continuum
limit of a heterostructure composed of a very large number of
dissimilar layers, leading to a smoothly varying effective mass.
As discussed earlier, similar spatial variations of the effective mass can also be engineered in ultracold-atom systems, where the tunneling amplitude in an optical lattice depends sensitively on the local lattice depth and can be
controlled in space using shaped optical potentials or digital
micromirror devices~\cite{Gauthier2016, Qiu2020, Wei2024, Tabares2025}. Moreover, the latest technological advances allow to probe ultracold Fermi gases directly in the continuum, making the Schr\"odinger equation~\eqref{eq:SE-slug0i} a potentially realizable setup, for both many-body Fermi systems~\cite{dejong2025, dixmerias2025} and at the level of a single particle~\cite{maltezopoulos2003, verstraten2025}.
The increase of the effective mass scales as
$m_\mathrm{eff}(x)= m_\mathrm{eff}\, |x|^{\alpha}$ with distance implies that the kinetic motion becomes
progressively suppressed at large distances. In other words, the particle becomes ``heavy'' as it moves further from the origin, thereby making it sluggish at larger distances.

A natural and interesting
question, therefore, is how this form of \emph{sluggish quantum mechanics}
(for $\alpha>0$) modifies the plane-wave eigenfunctions $e^{ikx}$ of a
conventional free quantum particle having a constant mass. 
Moreover, in the presence of an external confining potential $V_\mathrm{ext}(x)$, the Schr\"odinger equation in~\eref{eq:SE-slug0i} modifies to 
\begin{equation}
  i\hbar \frac{\partial \psi (x,t)}{\partial t}    =- \frac{\partial}{\partial x} \left(D(x) \frac{\partial \psi (x,t)}{\partial x}  \right)  + V_\mathrm{ext}(x) \psi(x,t),
  \label{eq:SE-slug-mod}
\end{equation}
with $D(x)=\hbar^2/[2m_\mathrm{eff}(x)]$.
For the $\alpha=0$ case, the Schr\"odinger equation~\eqref{eq:SE-slug-mod} with $D(x)=D_0=\hbar^2/(2m_\mathrm{eff})$, 
reduces to the
standard Schr\"odinger equation discussed in textbooks~\cite{landau2013quantum, albeverio2012solvable}. In particular, the harmonic potential $V_\mathrm{ext}(x)=\frac{1}{2}m_\mathrm{eff}\,\omega^2 x^2$ has found relevance in a wide range of contexts, both theoretically and experimentally. 
Therefore, it is natural
to ask whether there exists a class of external potentials $V_{\rm ext}(x)$ for which the Schr\"odinger equation~\eqref{eq:SE-slug-mod} admits exact analytical solutions for $\alpha>0$. Finally, for the $\alpha=0$ case, noninteracting spinless fermions in a harmonic potential constitute a fundamental and an exactly solvable many-body system whose ground-state and the kernel are deeply connected to the random matrix theory (RMT)~\cite{Vicari2012, Eisler2013,Marino2014, Calebrese2015, Dean2015, Dean2015b, Dean2016,Dean2019review}. In particular, the joint probability density function of the positions of the fermions coincides with that of the eigenvalues of the Gaussian unitary ensemble, leading to the Wigner semicircle law for the density profile in the thermodynamic limit~\cite{mehta2004random, forrester2010log}. It is, therefore, natural to address these questions for noninteracting spinless sluggish fermions in a trap for $\alpha>0$.

Before proceeding further, we would like to make some important comments on a classical counterpart of the quantum mechanical problem considered here. A closely related notion of sluggish dynamics has been explored in classical stochastic systems~\cite{SL2003,Cherstvy_2013,SCT_PRE,LB2029,SR22,AZ2023,SCT2023,Vecchio_2025,2025proxitaxis,boyer2026}, particularly, in the context of diffusion in heterogeneous media~\cite{Lenzi_Physica,MA2024}, where the mobility or the diffusivity or both are space-dependent. In such classical inhomogeneous systems, several variants of the diffusion equation are possible. Moreover, the diffusion operator need not be self-adjoint, unlike its quantum counterpart, which must be hermitian.  Two common examples of diffusion equations in inhomogeneous media are~\cite{vankampen2007stochastic} $\partial_t P(x,t) = \partial_x\left[D(x)\partial_x P(x,t)\right]$ and $\partial_t P(x,t) = \partial_x^2\left[D(x) P(x,t)\right]$. The former equation with $D(x)\sim |x|^{-\alpha}$ is the classical analogue of the quantum mechanical problem studied here. This class of models has been used to describe transport in disordered systems, crowded environments, and heterogeneous materials, where particles effectively become less mobile as they move away from certain regions. The quantum evolution equation in Eq.~\eqref{eq:SE-slug0i} may be viewed as a natural quantized analogue of such classical sluggish diffusion described by $\partial_t P(x,t) = \partial_x\left[D(x)\partial_x P(x,t)\right]$, with the role of the diffusion coefficient played by $D(x)=\hbar^2/[2m_\mathrm{eff}(x)]$. In this sense, our work provides a quantum generalization of the classical sluggish dynamics, enabling the exploration of quantum interference, spectral properties (eigenvalues and eigenvectors), and many-body quantum effects in systems with position-dependent diffusivity.

The main results of our paper are summarized as follows.
\begin{enumerate}[
    leftmargin=0pt,
    itemindent=2.5em,
    label=(\arabic*),
    labelsep=0.5em,
    itemsep=0pt,
    parsep=0pt,
    topsep=0pt,
    partopsep=0pt
]
\item We find the eigenfunctions and the quantum propagator for the sluggish quantum particle in one dimension, in the absence of an external potential, for the Schr\"odinger equation given by~\eref{eq:SE-slug0i}. 

\item We find the eigenfunctions $\psi_k(x)$ and the propagator for the Schr\"odinger equation~\eqref{eq:SE-slug-mod}, in the presence of the external potential $V_\mathrm{ext} = \frac{1}{2}m_\mathrm{eff}\,\omega^2 |x|^{\alpha+2}$.

\item We consider the many-body ground state of $N$ spinless noninteracting fermions in the external potential $V_\mathrm{ext} = \frac{1}{2}m_\mathrm{eff}\,\omega^2  |x|^{\alpha+2}$. 
The many-body ground-state wavefunction is given by the Slater determinant of the single particle eigenfunctions $\psi_k(x)$. 
Consequently, the joint probability density function of the positions of $N$ fermions can be written as a determinant $\frac{1}{N!} \det_{1\leq \,j,\,l\, \leq N} K_N(x_j, x_l)$, of the so-called quantum kernel $K_N(x,y)=\sum_{k=0}^{N-1}  \psi^*_k(x) \psi_k(y)$. In fact, these fermions form a `determinantal point process' where any $n$-point spatial correlation function can also be written as a determinant  $\det_{1 \leq \, j,\, l \, \leq n} K_N(x_j,x_l) $ of the kernel.
We obtain $K_N(x,y)$ exactly for the full-space $x\in(-\infty,\infty)$. Subsequently, we obtain the density profile in the large $N$ limit. We also analyze the scaling behavior of the kernel in the bulk, edge, and near the origin. We show that the scaled kernel near the origin $x=0$ for $\alpha>0$ is neither the Bessel nor the Airy kernel (that are standard for trapped fermions), but is new, and is given by a sum of two Bessel kernels with different indices.

\item We explicitly obtain the many-body ground-state wavefunction, and consequently,  the joint probability density function (JPDF) of the position of the $N$ fermions, for the half-space $x\in (0,\infty)$ with both Dirichlet and Neumann boundary conditions at $x=0$.
 \end{enumerate}   
    
The rest of the paper is organized as follows. In~\sref{s:derivation}, we derive the sluggish Schrodinger equation~\eqref{eq:SE-slug0i} by taking an appropriate continuum limit. We then solve the Schr\"odinger equation~\eqref{eq:SE-slug0i} to obtain its eigenfunctions and the full propagator in~\sref{s:sluggish-free}. In~\sref{s:sluggish-potential}, we obtain the eigenfunctions and also the propagator of the Schr\"odinger equation~\eqref{eq:SE-slug-mod} in the presence of the potential $V_\mathrm{ext}(x)=\frac{1}{2}m_\mathrm{eff}\,\omega^2|x|^{\alpha+2}$. In \sref{s:mapping}, we discuss a mapping that transforms the sluggish Schr\"odinger equation to the standard one with an effective potential given as a sum of harmonic and inverse square potential. 
Before discussing the many-body ground state of the sluggish quantum problem in an external confinement,  we first recapitulate the ground-state properties of $N$ noninteracting spinless fermions in a harmonic potential with $\alpha=0$ in~\sref{s:recap}. 
\Sref{s:mbgs} is devoted to studying the many-body ground state of $N$ noninteracting spinless sluggish fermions in an external trap $V_\mathrm{ext}(x)=\frac{1}{2}m_\mathrm{eff}\,\omega |x|^{\alpha+2}$. In \sref{s:JPDF-kernel}, we discuss the joint probability density function of the positions and the quantum correlation kernel. Using the kernel, we obtain the density profile in \sref{s:density}. \Sref{s:kernel} is devoted to the kernel and its scaling behavior in the bulk, edge, and near the origin.  We obtain the many-body wavefunction in the half-space $x\in (0,\infty)$ explicitly in \sref{s:JPDF-half}.  Finally, we conclude with an outlook in \sref{s:conclusion}. Certain details are relegated to the appendix.

\section{Derivation of the Sluggish Schr\"odinger equation}
\label{s:derivation}

The Schr\"odinger equation for the evolution of
any quantum state $\ket{\psi(t)}$ is given by
\begin{equation}
  i \hbar \frac{\partial}{\partial t}  \ket{\psi(t)} = H \ket{\psi(t)}
  \label{eq:SE1}
\end{equation}
In the position basis $\ket{x}$, the Hamiltonian for a single quantum particle in a potential $V(x)$ is diagonal and is given by
\begin{equation}
    H_x \equiv \braket {x|H|x}= -\frac{\hbar^2}{2 m}\frac{\partial^2}{\partial x^2} + V(x)\, .
    \label{eq:Hx}
\end{equation}
Therefore, the wavefunction $\psi(x,t)=\braket{x|\psi(t)}$ of a quantum particle in an external potential $V(x)$ becomes
\begin{equation}
    i\hbar \frac{\partial \psi(x,t)}{\partial t} = H_x \psi(x,t)\,.
    \label{eq:SE}
\end{equation}

Consider that the external potential $V(x)$ consists of a sequence of deep potential wells, whose minima $\{x_i\}$ are separated by some characteristic length scale $\delta$. In this case, the quantum particle is mostly localized around the minima, and therefore, 
it is more convenient to describe the Schr\"odinger equation~\eqref{eq:SE1} in an effective localized basis $\ket{x_i}$, with the Hamiltonian
\begin{math}
    H=\sum_{i,j} \ket{x_i} \braket{x_i|H|x_j} \bra{x_j}.
\end{math}
In the nearest neighbor tight-binding approximation, $H_{i,j} \equiv \braket{x_i|H|x_j}$ is given by 
\begin{equation}
  H_{i,j} = \begin{cases}
      0 &\text{for }~|x_i-x_j| >\delta,\\
      a(x_i, x_i\pm\delta) & \text{for }~ x_j=x_i\pm\delta ,\\
      b(x_i, \delta) &\text{for }~ i=j  
  \end{cases}  
  \label{eq:Hij}
\end{equation}
with  $H_{j,i}=H_{i,j}$.
The hopping elements can, in principle, be computed from
\begin{equation}
    a(x_i, x_i+\delta) = \int_{-\infty}^\infty dx\, \phi^*(x_i, x) H_x \phi(x_i+\delta, x)\, ,
\end{equation}
and
\begin{equation}
    b(x_i,\delta)=\int_{-\infty}^\infty dx\, \phi^*(x_i, x) H_x \phi(x_i, x)\, ,
\end{equation}
where $H_x$ is given in \eref{eq:Hx} and
$\phi(x_i,x)\equiv \braket{x|x_i}$ is a localized function around the minimum $x_i$, which depends on $\delta$ and the depth of the trap. For a periodic potential $V(x)$, the localized wavefunctions $\phi(x_i,x)$ are the so-called Wannier functions~\cite{marder2010condensed}.

The time-dependent Schr\"odinger equation~\eqref{eq:SE1} 
in the localized basis $\ket{x_i}$ can be written as
\begin{equation}
    i \hbar \frac{\partial \psi_\text{tb}(x_i,t)}{\partial t}  = \sum_j H_{i,j} \psi_\text{tb}(x_j,t)\,,
\label{eq:SE-tb}
\end{equation}
where the subscript `tb' stands for tight-binding and $\psi_\text{tb}(x_i,t)=\braket{x_i|\psi(t)}$ is the projection of the state $\ket{\psi(t)}$ to the localized basis and $H_{i,j}$ given in \eref{eq:Hij}. It is easy to see that the position-basis wavefunction $\psi(x,t)$ and the tight-binding wavefunction wavefunction $\psi_\text{tb}(x_i,t)$ are related by $\psi(x,t) =  \sum_i \psi_\text{tb}(x_i,t)\, \phi(x_i,x)$.

We now assume that the hopping elements have the form
\begin{equation}
    a(x_i,x_i\pm \delta) = -\frac{J}{2} \bigl[D(x_i)+D(x_i\pm \delta)\bigr]
    \label{eq:a}
\end{equation}
and the onsite term
\begin{equation}
    b(x_i, \delta)= -\bigl[a(x_i,x_i+\delta) + a(x_i,x_i-\delta)\bigr]\, .
    \label{eq:b}
\end{equation}

The tight-binding Hamiltonian in \eqref{eq:Hij}, can be written  as
\begin{align}
    H=\sum_i \Bigl[&a(x_i+\delta, x_i) \ket{x_i}  \bra{x_i+\delta} + a(x_i, x_i+\delta)\ket{x_i+\delta} \bra{x_i} \notag\\ &- \bigl(a(x_i, x_i+\delta)+ a(x_i,x_i-\delta)\bigr)\ket{x_i}\bra{x_i}\Bigr]\,.
\label{eq:Hamiltonian0}
\end{align}
Therefore, the Schr\"odinger equation ~\eqref{eq:SE-tb} of the tight-binding model becomes, 
\begin{align}
    i\hbar \frac{\partial \psi_\text{tb}(x_i,t)}{\partial t}   &= a(x_i,x_i+\delta)\psi_\text{tb}(x_i+\delta,t)\notag\\ &+ a(x_i,x_i-\delta)\psi_\text{tb}(x_i-\delta,t) \cr &- \bigl[a(x_i,x_i+\delta)+ a(x_i,x_i-\delta)\bigr] \psi_\text{tb}(x_i,t).
    \label{eq:SE-tb2}
\end{align}

We now take the continuum limit of the tight-binding Schr\"odinger
equation~\eqref{eq:SE-tb2} by letting the lattice spacing $\delta \to 0$.
Expanding the right-hand side of Eq.~\eqref{eq:SE-tb2} in a Taylor series
in $\delta$, one finds that the terms of order $O(\delta^0)$ and
$O(\delta)$ cancel identically, and the leading nonvanishing contribution
appears at order $O(\delta^2)$.

Using the expression for $a(x_i,x_i+\delta)$ from~\eref{eq:a}, together
with 
\begin{equation}
\begin{split}
    a(x_i,x_i+\delta)\big|_{\delta\to 0}  &= - J D(x_i),\\
    \left[\frac{\partial}{\partial y} a(x_i,y) \right]_{y\to x_i}
    &= -\frac{J}{2} D'(x_i),
\end{split}
\label{eq:a_lim}
\end{equation}
the Taylor expansion of Eq.~\eqref{eq:SE-tb2} yields 
\begin{align}
    \frac{\partial \psi_\text{tb}(x_i,t)}{\partial t}
    = &-\delta^2 J \left[
    D'(x_i)\, \frac{\partial \psi_\text{tb}(x_i,t)}{\partial x_i}
    + D(x_i)\, \frac{\partial^2 \psi_\text{tb}(x_i,t)}{\partial x_i^2}
    \right]\notag\\
    &+ J\, o(\delta^2),
    \label{eq:SE-conti}
\end{align}
where $o(\delta^2)$ denotes terms of higher order than $O(\delta^2)$.

In the limit $\delta \to 0$ and $J \to \infty$, keeping
$\delta^2 J = 1$ fixed, the higher-order terms $J\,o(\delta^2)$
vanish, and we get the following effective Schr\"odinger equation
\begin{equation}
  i\hbar \frac{\partial \psi_\text{tb}(x_i,t) }{\partial t}   =- \frac{\partial}{\partial x_i} \left(D(x_i) \frac{\partial \psi_\text{tb} (x_i,t)}{\partial x_i}  \right)  .
  \label{eq:SE-slug0}
\end{equation}

In this continuum limit,  $x_i$ in \eref{eq:SE-slug0} becomes a continuous variable.  With a slight abuse of notation, we drop the subscript $i$ from $x_i$ and  replace it with
  $x$. We also omit the subscript `tb' in $\psi_\text{tb}$ and denote $\psi_\text{tb}(x_i,t)$ simply by $\psi(x,t)$, and rewrite \eref{eq:SE-slug0} in the standard notation as
\begin{equation}
   i\hbar \frac{\partial \psi(x,t) }{\partial t}   =- \frac{\partial}{\partial x} \left(D(x) \frac{\partial \psi (x,t)}{\partial x}  \right)  .
  \label{eq:SE-slug}  
\end{equation} 
\Eref{eq:SE-slug} is identical to \eref{eq:SE-slug0i}, with the identification $D(x)=\hbar^2/[2m_\mathrm{eff}(x)]$.
In addition to the sequence of potential wells $V(x)$, if there is a slowly varying confining potential $V_\mathrm{ext}(x)$ in \eqref{eq:SE}, the resulting sluggish Schr\"odinger equation yields \eref{eq:SE-slug-mod}.

We emphasize that the variable
$x$ in Eqs.~\eqref{eq:SE-slug0i} and \eqref{eq:SE-slug-mod}, emerges from the coarse-grained lattice description and should not be identified with the microscopic spatial variable appearing in the
original Schr\"odinger equation~\eqref{eq:SE}. Although we denote
both by the same symbol to cast Eqs.~\eqref{eq:SE-slug0i} and \eqref{eq:SE-slug-mod} in the standard well-known form, these equations and \eref{eq:SE}  belong to two different levels of description. For the rest of the paper, the function $\psi(x,t)$ should be identified with the continuum limit of the tight-binding wavefunction $\psi_\text{tb}(x_i,t)$, and not with the position-basis wavefunction in \eref{eq:SE}.
 In this paper, we consider a position-dependent coefficient of the form
$D(x)=D_0/|x|^{\alpha}$, with $\alpha>0$.

For the special case $\alpha=0$,
\eref{eq:SE-slug} reduces to the Schr\"odinger equation for a free particle with an effective mass $m_\mathrm{eff}$,
i.e.,~\eref{eq:SE} with $V(x)=0$, upon identifying
$D_0=\hbar^2/(2m_\mathrm{eff})$. In other words, for a periodic potential $V(x)$ whose wells
have uniform depth in space---so that both the hopping amplitudes and the onsite energies in the corresponding
tight-binding model are spatially independent---the continuum limit of the
tight-binding Schr\"odinger equation reduces to that of a free particle with an effective mass $m_\mathrm{eff}$~\cite{feynman2015feynman}. For $\alpha>0$, as mentioned earlier, the quantum particle becomes ``heavy'' at larger distances, and the quantum dynamics becomes sluggish.

In the next section (\sref{s:sluggish-free}), we solve for eigenfunctions of the sluggish Schr\"odinger equation~\eqref{eq:SE-slug0i}, i.e., the free case [which is the same as  \eref{eq:SE-slug}],  and use them to obtain the propagator. 

\section{Sluggish quantum particle without external potential}
\label{s:sluggish-free}

The general solution of the sluggish Schr\"odinger equation~\eqref{eq:SE-slug} [which is same as \eref{eq:SE-slug0i}] can be written as
\begin{equation}
    \psi(x,t) = \int_0^\infty dE\, C(E)\, e^{-(i/\hbar)\, E t}\, f_E(x)\, .
    \label{eq:psi1}
\end{equation}
The coefficients $C(E)$ in \eref{eq:psi1}
are fixed by the initial condition $\psi(x,0)$
and $f_E(x)$ satisfies the eigenvalue equation, i.e.,  the time-independent Schr\"odinger equation,
\begin{equation}
    \frac{d}{dx} \left[D(x)\frac{df_E(x)}{dx}\right] + E f_E(x)=0.
    \label{eq:SE-timeInd}
\end{equation}

We now consider the specific form $D(x)=D_0/|x|^\alpha$. Moreover, we first consider the case $x>0$. \Eref{eq:SE-timeInd} becomes
\begin{equation}
f''_\lambda(x) -\frac{\alpha}{x}f'_\lambda(x) + \lambda x^\alpha f_\lambda(x) =0 ,
\label{eq:f-diff}
\end{equation}
where $\lambda=E/D_0$.
Let us substitute the anzatz 
\begin{equation}
    f_\lambda(x) = x^\kappa g(a x^\beta)
    \label{eq:flambda}
\end{equation}
in \eref{eq:f-diff}, with arbitrary $a$, $\beta$, and $\kappa$. This yields
\begin{align}
    &g''(a x^\beta) +\frac{2\kappa+\beta-\alpha-1}{\beta a x^\beta} g'(a x^\beta) \cr &+ \frac{\lambda x^{\alpha+2}}{(\beta a x^\beta)^2}\left[1-\frac{\kappa(1+\alpha)-\kappa^2}{\lambda x^{\alpha+2}}\right] g(a x^\beta) =0\, .
\end{align}
Now choosing 
\begin{equation}
  \kappa=\frac{\alpha+1}{2},\quad \beta=\frac{\alpha+2}{2}, \quad a=\frac{2\sqrt{\lambda}}{\alpha+2},  
\end{equation} 
and defining 
\begin{equation}\label{eq:nu_nutilde}
    \nu = \frac{1}{\alpha + 2} \quad \text{and}\quad \tilde \nu = \frac{\alpha + 1}{\alpha + 2} = 1-\nu\,,
\end{equation}
we find that $g(z)$ satisfies the Bessel differential equation~\cite{gradshteyn2014table} 
\begin{equation}
    g''(z)+ \frac{1}{z} g'(z) + \left(1-\frac{\tilde{\nu}^2}{z^2}\right) g(z) =0 \, .
    \label{eq:bessel}
\end{equation}
Since $\tilde{\nu}\in[1/2,1)$ is non-integer, the two linearly independent solutions of \eref{eq:bessel} can be chosen to be the two Bessel functions $J_{\tilde{\nu}}(z)$  and $J_{-\tilde{\nu}}(z)$. Therefore, from \eref{eq:flambda}, the two sets of eigenfunctions of 
\eref{eq:f-diff}, for $x>0$,  are given by
\begin{align}
\label{eq:f1}
    f_\lambda^{+} (x) & = x^{\frac{\tilde \nu}{2\nu}}J_{ -\tilde \nu}(2\nu \sqrt{\lambda}x^{\frac{1}{2\nu}}), \\
     \label{eq:f2}
     f_\lambda^{-} (x) &= x^{\frac{\tilde \nu}{2\nu}}J_{\tilde \nu}(2\nu \sqrt{\lambda}x^{\frac{1}{2\nu}}),
\end{align}
and the general solution of \eref{eq:f-diff} in the half-space $x\in(0,\infty)$, is given by the linear combination 
\begin{equation}
f_\lambda(x) =  C_+(\lambda) \,f_\lambda^{+} (x) + C_{-}(\lambda) \, f_\lambda^{-} (x).
\label{eq:f-sol}
\end{equation}

Since for $D(x)=D_0/|x|^\alpha$, \eref{eq:SE-timeInd} is symmetric under $x\to -x$,  there are two types of eigenfunctions: one set is symmetric under $x\to -x$, i.e., $f_\lambda(x) = f_\lambda(-x)$, while the other is anti-symmetric, i.e., $f_\lambda(x) = - f_\lambda(-x)$. We can construct $f_\lambda(x)$ for $x<0$ from \eref{eq:f-sol} by properly identifying the symmetric and the anti-symmetric eigenfunctions. The anti-symmetric wavefunctions must vanish as $x \to 0^+$, whereas the symmetric ones must go to a non-zero constant, so that their derivative vanishes at $x=0^+$.

Using $J_{\pm\tilde{\nu}}(z) \sim z^{\pm \tilde{\nu} }$, it is easy to see that $f_\lambda^{-}(x)\to 0$ as $x\to 0^+$, whereas $f_\lambda^{+}(x)$ tends to a constant. Therefore, $f_\lambda^{+}(x)$ belongs to the even set of solutions and $f_\lambda^{-}(x)$ are the odd ones. Therefore, for $x<0$, the even and odd sets of eigenfunctions are given by $f_\lambda^{+}(-x)$ and $-f_\lambda^{-}(-x)$ respectively. 

Any initial state $\psi(x,0)\equiv\psi_0(x)$ can be decomposed into a symmetric
and an anti-symmetric component as
\begin{equation}
\psi_0(x)=\psi_0^+(x) + \psi_0^-(x)\quad \text{with}\quad\psi_0^\pm (x) = \frac{\psi_0(x)\pm \psi_0(-x)}{2}.
\end{equation}
Each component evolves independently under the corresponding propagator $G^\pm(x,t|x_0,0)$ as
\begin{equation}
   \psi^\pm (x,t) =\int_0^\infty\, dx_0\, \psi_0^\pm (x_0)\, G^\pm(x,t|x_0,0)\,,\quad\text{for ~} x>0. 
   \label{eq:psi_pm}
\end{equation}
We define the even and the odd quantum propagators by~\eref{eq:psi_pm}, which are to be determined.  The symmetric (even) and the anti-symmetric (odd) wavefunctions for $x<0$ can be obtained from the symmetry $\psi^+(-x,t)=\psi^+(x,t)$ and $\psi^-(-x,t)=-\psi^-(x,t)$. The complete wavefunciton is simply given by $\psi(x,t)=\psi^+(x,t) + \psi^-(x,t)$. Therefore, it suffices to focus only on the positive domain $x>0$, and our goal now is to obtain the symmetric and anti-symmetric propagators $G^\pm(x,t|x_0,0)$ for $x>0$ and $x_0>0$ in \eref{eq:psi_pm}.

For the symmetric and the anti-symmetric initial states $\psi^\pm(x_0)$, respectively, only 
the symmetric and the anti-symmetric eigenfunctions $f_\lambda^{\pm}(x)$, respectively, contribute to the general solution~\eqref{eq:psi1}. Therefore, for $x>0$, we have
\begin{equation}
    \psi^\pm(x,t) = \int_0^\infty dE\, C_\pm (E) \, e^{-(i/\hbar)\, E t}\, f_\lambda^{\pm}(x),
    \label{eq:psi-pm2}
\end{equation}
where $f_\lambda^{\pm}(x)$ are given in Eqs.~\eqref{eq:f1} and \eqref{eq:f2} with $\lambda=E/D_0$. The constants $C_\pm(E)$ are determined by the initial condition
\begin{equation}
    \psi_0^\pm(x)= \int_0^\infty dE\, C(E)\, f_\lambda^{\pm}(x).
    \label{eq:psi0}
\end{equation}
Multiplying both sides of \eref{eq:psi0} by $f_{\lambda'}^\pm(x)$, respectively, then integrating over $x$ from $0$ to $\infty$, and using the orthogonality of the Bessel functions 
\begin{equation}
    \int_0^\infty z \,J_{\nu} (kz)\, J_\nu (k' z)\, dz=\frac{1}{k}\delta(k-k'),
    \label{eq:bessel-ortho}
\end{equation}
we get
\begin{equation}
    C_\pm(E) = (\nu/D_0)\int_0^\infty dx\, \psi_0^\pm (x)\, f_{\lambda}^{\pm}(x).
    \label{eq:CE-pm}
\end{equation}
Therefore,  substituting $C_\pm(E)$ from \eref{eq:CE-pm} in \eref{eq:psi-pm2}, we see that the
symmetric and the anti-symmetric components of the wavefunctions $\psi^\pm(x,t)$, at any time $t$, are given by \eref{eq:psi_pm}, where the the symmetric and anti-symmetric quantum propagators $G^\pm(x,t|x_0,0)$, for $x>0$ and $x_0>0$, are given by
\begin{align}
    G^\pm(x,t|x_0,0)&= \nu\, (xx_0)^{\frac{\tilde \nu}{2\nu}}\int_0^\infty d\lambda\,  e^{-(iD_0/\hbar)\, \lambda t}\cr &\times \,J_{\mp\tilde \nu}\left(2\nu \sqrt{\lambda}\,x_0^{\frac{1}{2\nu}}\right)\, J_{\mp\tilde \nu}\left(2\nu \sqrt{\lambda}\, x^{\frac{1}{2\nu}}\right).
    \label{eq:propagator1}
  \end{align}
Now, using the following integral~\cite{gradshteyn2014table}
\begin{align}
    \int_0^\infty &e^{-a \lambda}  J_{\pm\tilde{\nu}} \left(2b \sqrt{\lambda}\right)\, J_{\pm\tilde{\nu}} \left(2c \sqrt{\lambda}\right)\, d\lambda \cr &= \frac{1}{a} I_{\pm \tilde{\nu}} \left(\frac{2bc}{a}\right) \exp\left(-\frac{b^2+c^2}{a}\right),
\end{align}
the integral in \eref{eq:propagator1}  over $\lambda$ can be carried out, yielding
\begin{align}
    G^\pm(x,t|x_0,0)=& -\frac{i \hbar \nu}{D_0 \,t}\, i^{\pm \tilde{\nu}}\, (xx_0)^{\frac{\tilde \nu}{2\nu}}\, J_{\mp\tilde{\nu}} \left(\frac{2\hbar \nu^2}{D_0\, t} (xx_0)^{\frac{1}{2\nu}}\right)\notag \\& \times\, \exp\left[\frac{i\hbar \nu^2}{D_0\, t} \left(x^{1/\nu}+ x_0^{1/\nu}\right)\right], 
    \label{eq:propagator2}
\end{align}
where we have used the relation $I_{\mp\tilde{\nu}}(-i z)= i^{\pm\tilde{\nu}}J_{\mp\tilde{\nu}}(z) $. Using \eref{eq:propagator1} and the orthogonality relation \eqref{eq:bessel-ortho}, it can be shown that 
\begin{equation}
    \int_0^\infty G^\pm (x,t|x_0,0) \, G^{\pm *} (x,t|x_0',0)\, dx =\delta(x_0-x_0').
    \label{eq:prop-ortho}
\end{equation}
Moreover, using \eref{eq:prop-ortho} in \eref{eq:psi_pm}, we can easily verify the conservation of the probability $\int_0^\infty |\psi^\pm(x,t)|^2\, dx = \int_0^\infty|\psi_0^\pm(x_0)|^2\, dx_0 $, at all times, as expected.

Let us verify that, for $\alpha=0$, \eref{eq:propagator2} reduces to the well-known result for a conventional free quantum particle. For $\alpha=0$, from \eref{eq:nu_nutilde}, we have  $\nu=\tilde{\nu}=1/2$. Using $J_{-1/2}(z) =\sqrt{2}\, \cos(z)/\sqrt{\pi x}$ and $J_{1/2}(z) =\sqrt{2}\, \sin(z)/\sqrt{\pi x}$ in \eref{eq:propagator2}, after some straightforward algebra, we get 
\begin{align}
G^\pm(x,t|x_0,0)=\frac{\sqrt{\hbar}}{\sqrt{4 i \pi D_0 \,t}} \Biggl[&\exp\left(\frac{i\hbar}{4 D_0 \,t} (x-x_0)^2\right) \notag \\ \pm &\exp\left(\frac{i\hbar}{4 D_0 \,t} (x+x_0)^2\right)\Biggr].
\label{eq:g-pm-free}
\end{align}
Therefore, in the special case $\alpha=0$, we recover the well-known symmetric and anti-symmetric propagators of a free quantum particle, with $D_0=\hbar^2/(2 m_\mathrm{eff})$.

To summarize, \eref{eq:g-pm-free} provides the propagators needed to obtain the symmetric and the anti-symmetric components of the wavefunctions from \eref{eq:psi_pm} for $x>0$. The corresponding wavefunctions for the negative side are then obtained simply by using the symmetry $\psi^+(-x,t)=\psi^+(x,t)$ and $\psi^-(-x,t)=-\psi^-(x,t)$. Finally, the complete wavefunction is  given by $\psi(x,t)=\psi^+(x,t) + \psi^-(x,t)$. Alternatively, we can write down the quantum propagator on the full space $x\in(-\infty, \infty)$ as
\begin{align}
    G(x,t|x_0,0)= \frac{1}{2}\Bigl[& G^+\bigl(|x|,t\big||x_0|,0\bigr) \notag\\ & + \mathrm{sgn}(x)\, \mathrm{sgn}(x_0)\, G^-\bigl(|x|,t\big||x_0|,0\bigr)\Bigr],
\end{align}
and the wavefunction for any initial state $\psi_0(x)$ is now given by
\begin{equation}
    \psi(x,t) = \int_{-\infty}^\infty\, dx_0\, \psi_0(x_0)\, G(x,t|x_0,0).
\end{equation}

Having solved the problem in the absence of a trap, we next study the single particle problem in an external trap $V_\text{ext}(x) = \frac{1}{2}m_\text{eff}\omega^2 |x|^{\alpha+2}$.

\section{Sluggish Single Particle Schr\"odonger equation with an external potential}
\label{s:sluggish-potential}
Now we consider the sluggish quantum particle in an external potential $V_{\rm ext}(x)$. The general solution of the Schr\"odinger equation~\eqref{eq:SE-slug-mod} is given by
\begin{equation}
    \psi(x,t)= \int_0^\infty dE\, C(E)\, e^{-(i/\hbar) E t} f_E(x),
    \label{eq:psixt-pot}
\end{equation}
where $f_E(x)$ denote the eigenfunction associated with energy $E$ and the coefficient $C(E)$ is determined from the initial conditions $\psi(x,0)$. The eigenfunction $f_E(x)$ satisfies the time-independent Schr\"odinger equation
\begin{equation}
    \frac{d}{dx} \left[D(x)\frac{df_E(x)}{dx}\right] + \bigl[E - V_\mathrm{ext}(x)\bigr]\, f_E(x)=0.
    \label{eq:SE-timeInd-pot}
\end{equation}
We recall here that $D(x) = D_0/|x|^\alpha$ with $\alpha>0$. We will henceforth focus on external potentials that are symmetric around $x=0$. For such symmetric potentials,
there are two sets of eigenfunctions, namely, even (symmetric) and odd (anti-symmetric) under $x\to -x$. Therefore, it suffices to find the eigenfunction for $x>0$, from which the eigenfunctions for $x<0$ can be constructed by symmetry considerations, as done in \sref{s:sluggish-free}. The even eigenfunctions satisfy the Neumann boundary condition at $x=0$, i.e., $f_E'(0)=0$. In this case, the even eigenfunction has a non-zero value at $x=0$, i.e., $f_E(0)\neq 0$. In contrast, the odd eigenfunctions satisfy the Dirichlet boundary condition at $x=0$, i.e., $f_{E}(0)=0$.

Let us briefly recall the case $\alpha=0$, where Eq.~\eqref{eq:SE-timeInd-pot} is the standard Schr\"odinger equation with an external potential $V_{\rm ext}(x)$. Even in this case, the Schr\"odinger equation is not generally solvable for arbitrary $V_{\rm ext}(x)$. However, a special role is played by the harmonic potential $V_{\rm ext}(x) =\frac{1}{2} m_{\rm eff} \omega^2 x^2$ with $m_{\rm eff}$ being a constant. The harmonic potential for $\alpha=0$ is special because i) the explicit eigenfunctions are known in terms of Hermite polynomials and Gaussians (see below) and ii) it is easily realized as a trapping potential in typical cold-atom experiments~\cite{exp-harmonic}. 
We recall that for the harmonic potential, the allowed energies are discrete and given by $E_k = (k+1/2) \hbar \omega$, where $k=0,1,\dots,\infty$, and the associated eigenfunctions $f_{E_k}(x)$ are given by
\begin{equation}
 \label{eq:Hermite1}
f_{E_k}(x) = \left[ \frac{\sqrt{\mu}}{\sqrt{\pi} 2^k k!}\right]^{1/2} \, e^{-\frac{\mu\,x^2}{2}} H_k(\sqrt{\mu} \,x) \;,
\end{equation}
 where
 \begin{equation}
     \mu=\frac{\sqrt{m_\mathrm{eff}}\,\omega}{\sqrt{2 D_0}} \equiv \frac{m_\mathrm{eff}\,\omega}{\hbar},
     \label{eq:mu-def}
 \end{equation}
and  $H_k(z)$ is the  Hermite polynomial of degree $k$. When $k$ is even, the derivative of the eigenfunction vanishes at $x=0$ (Neumann). This corresponds to the even sector. In contrast, when $k$ is odd, the eigenfunction $f_{E_k}(x)$ vanishes at $x=0$ (Dirichlet). This is the odd sector. 

For $\alpha>0$, it is then natural to ask if there in analogous solvable potential which is a cousin of the harmonic one for $\alpha=0$. Since for $\alpha=0$ the harmonic potential is given by $V_{\rm ext}(x) = \frac{1}{2}m_{\rm eff} \omega^2 x^2$, it is then natural to consider a potential of the form $V_{\rm ext}(x) = \frac{1}{2}m_{\rm eff}(x) \omega^2 x^2$, where we recall that $m_{\rm eff}(x) = m_{\rm eff}|x|^{\alpha+2}$ with $m_{\rm eff}$ a constant, as discussed in the introduction.
Thus, for $\alpha>0$, one expects that the natural potential that extends the harmonic ($\alpha=0$) case is
 \begin{equation}
    V_\mathrm{ext}(x) = \frac{1}{2}m_\mathrm{eff}(x)\omega^2 x^2 =\frac{1}{2}m_\mathrm{eff}\, \omega^2 |x|^{\alpha+2}. 
    \label{eq:V-eff}
 \end{equation}
Indeed, we show below that the Schr\"odinger equation Eq.~\eqref{eq:SE-timeInd-pot} is exactly solvable for the external  potential given in \eref{eq:V-eff}. 

To find the eigenfunctions of Eq.~\eqref{eq:SE-timeInd-pot} with $V_{\rm ext}(x)$ given in \eref{eq:V-eff}, it is convenient to focus on $x>0$ and exploit the symmetry of the odd and even sectors.
For $x>0$, \eref{eq:SE-timeInd-pot} can be explicitly written as
\begin{equation}
    f''_\lambda(x) -\frac{\alpha}{x}f'_\lambda(x) + \left[\lambda  - \mu^2\, x^{\alpha+2}\right]x^\alpha f_\lambda(x) =0 ,
\label{eq:f-diff-pot}
\end{equation}
where $\lambda=E/D_0$, $\mu=\sqrt{m_\mathrm{eff}}\,\omega/\sqrt{2 D_0} \equiv m_\mathrm{eff}\,\omega/\hbar$.
To proceed, we make the substitution 
\begin{equation}
    f_\lambda(x) = e^{-c \,x^\kappa} g(d\, x^\beta)
\label{eq:f-ans2}
\end{equation}
in \eref{eq:f-diff-pot} and obtain a differential equation for $g(z)$. Next, by choosing
\begin{equation}
    c=\frac{d}{2} = \frac{\mu}{\alpha+2}, \quad\text{and}\quad\kappa=\beta=\alpha+2,  
\end{equation}
we find that $g(z)$ satisfies the Kummer differential equation~\cite{gradshteyn2014table, abramowitz1965handbook} 
\begin{equation}
    z g''(z) + (b-z) g'(z) - a\, g(z)=0,
    \label{eq:kummer}
\end{equation}
with 
\begin{equation}
    b=\frac{1}{\alpha+2}\quad\text{and}\quad a= \frac{\mu-\lambda}{2 \mu (\alpha+2)}\, .
    \label{eq:ab}
\end{equation}

The two independent solutions of \eref{eq:kummer} can be taken as the Kummer's function of the first kind $M(a,b,z)$ and the Tricomi confluent hypergeometric function (Kummer's function of the second kind)
$U(a,b,z)$. The Kummer's function diverges as $M(a,b,z)\sim e^z z^{a-b}$  as $z\to \infty$~\cite{abramowitz1965handbook}. Consequently, using $g(z)=M(a,b,z)$ \eref{eq:f-ans2}, results in a divergence of $f_\lambda(x)$ as $x\to\infty$. On the other hand, $U(a,b,z) \sim z^{-a}$ as $z\to\infty$~\cite{abramowitz1965handbook}. Therefore, we discard the unphysical solution, and $f_\lambda(x)$ in \eref{eq:f-ans2} is given by (up to a multiplicative constant)
\begin{equation}
    f_\lambda(x) =  e^{-\frac{\mu x^{\alpha+2}} {\alpha+2}}\, U\left(a,b, \frac{2\mu\, x^{\alpha+2}}{\alpha+2}\right)\, .
    \label{eq:f-pot3}
\end{equation}
where $a$ and $b$ are given in \eref{eq:ab}. 

\subsection{Eigenvalues and eigenfunctions}

We now find the quantization condition for the eigenvalues $\lambda$. This is given by the boundary condition at $x=0$. Therefore, we investigate \eref{eq:f-pot3} as $x\to 0$. Let us  assume that $f_\lambda(x) \sim x^{\phi}$ as $x\to 0$ and substitute this in the differential equation~\eqref{eq:f-diff-pot}. 
This gives $
\phi(\phi-1-\alpha)=0$,
indicating that there are two families of eigenvalues parametrized by the two values of $\phi$, namely
$\phi=0$ and $\phi=1+\alpha$. For $\alpha=0$ (harmonic oscillator), these two are the analogues of the `even' and the `odd' eigenfunctions, respectively. Indeed, using series expansion of $U(a,b,z)$~\cite{abramowitz1965handbook}, the solution $f_\lambda(x)$ in \eqref{eq:f-pot3}, when expanded for small $x$, gives
the two families as 
\begin{align}
f_\lambda(x) &=  \frac{1}{\Gamma\left(1+a-b\right)}\, \left[a_0+ a_1\, x^{\alpha+2}+ a_2\, x^{2(\alpha+2)}+\ldots\right]\notag\\
&+ \frac{x^{\alpha+1}}{\Gamma\left(a\right)}\left[b_0+ b_1\, x^{\alpha+2} + b_2\,  x^{2 (\alpha+2)}+\ldots\right],
\label{smallx.1}
\end{align}
where  $\{a_0, a_1, a_2, \dotsc , b_0, b_1, b_2, \dotsc\}$ are known coefficients, and $a$ and $b$ are given in \eref{eq:ab}. The first series in \eref{smallx.1} corresponds to the even branch of the eigenfunctions as it tends to a constant as $x\to 0$. On the other hand, the second series in \eref{smallx.1} tends to zero as $x\to 0$, and therefore, corresponds to the odd branch of the eigenfunctions.

\subsubsection{Even eigenfunctions}

To select the family of eigenfunctions belonging to $\phi=0$, i.e., the `even' sector, we need to set $a=-n$ with $n=0,1,2, \dotsc,$ so that  $\Gamma(a)$ diverges, and consequently, the second term in \eqref{smallx.1} vanishes. Using the expression of $a$ from \eref{eq:ab}, gives  the quantized eigenvalues corresponding to the even eigenfunctions as
\begin{equation}
\lambda_n^\mathrm{even} = 2(\alpha+2)\mu\, n+ \mu, \quad {\rm where}\quad n=0,1,2, \ldots\, .
\label{eq:even-lambda}
\end{equation}
For convenience, we will denote the even eigenfunctions by $\phi_n(x)$ with $n=0,1,2, \dotsc$.  They are obtained by setting $\lambda=\lambda_n^\mathrm{even}$ and $a=-n$ in Eq.~\eqref{eq:f-pot3}. For integer $n$, we have~\cite{abramowitz1965handbook} 
\begin{equation}
\label{eq:UL}
U(-n, b, z) = (-1)^n n! \,L_n^{(b-1)} (z)\, ,
\end{equation}
where  
\begin{equation}
 L_n^{(\gamma)}(z)  = \sum_{i=0}^n (-1)^i \binom{n+\gamma}{n-i}\frac{z^i}{i!}
  \label{eq:Laguerre}   
\end{equation}
is the generalized  Laguerre polynomial (also known as the associated Laguerre polynomial)~\cite{abramowitz1965handbook}. Indeed,  $L_n^{(\gamma)}(z)$ is a polynomial solution of the differential equation \eqref{eq:f-diff-pot} with $a=-n$ and $b=\gamma +1$. Although \eref{eq:f-pot3} has been obtained for $x>0$, we can obtain the even eigenfunction for the whole space $x\in(-\infty, \infty)$ by making the eigenfunction symmetric. Therefore, we have
\begin{equation}
    \phi_n(x)= A_n\,
    e^{-\frac{\mu |x|^{\alpha+2}} {\alpha+2}}\, L_n^{(-\frac{\alpha+1}{\alpha+2})}\left( \frac{2\mu\, |x|^{\alpha+2}}{\alpha+2}\right)
    \label{eq:phin}.
\end{equation}
The constant $A_n$ can be determined using the normalization condition, $\int_{-\infty}^\infty \phi_m(x) \phi_n(x)\, dx = \delta_{m,n}$. Using the  orthogonality relation of the generalized Laguerre polynomials~\cite{abramowitz1965handbook}
\begin{equation}
    \int_0^\infty  z^\gamma \,e^{-z}\, L_n^{(\gamma)}(z)\, L_m^{(\gamma)} (z)\, dz = \frac{\Gamma(n+1+\gamma)}{n!}\, \delta_{n,m}\, ,
    \label{eq:Lag-ortho}
\end{equation}
we find
\begin{equation}
    A_n =  \sqrt{\left(\frac{2\mu}{\alpha+2}\right)^{-\gamma}\, \frac{\mu \, n!}{\Gamma(n+1-\gamma)}}\quad\text{with}\quad \gamma=\frac{\alpha+1}{\alpha+2}.
    \label{eq:Bn}
\end{equation}
If the eigenfunction is normalized to unity over the half-space $x\in (0,\infty)$, then the normalization constant $A_n$ becomes $\sqrt{2}$ times the expression of $A_n$ in \eqref{eq:Bn}. 

For $\alpha=0$, using $L_n^{(-1/2)}(x^2) = (-1)^n H_{2n}(x)/\bigl(n! 2^{2n}\bigr)$, with $H_{2n}(x)$ being the even Hermite polynomials~\cite{abramowitz1965handbook}, it is easy to check that $\phi_n(x)$ in \eqref{eq:phin} reduces to the even eigenfunctions of a quantum harmonic oscillators, given in \eref{eq:Hermite1} with $k=2n$.

\subsubsection{Odd eigenfunction}

To find the family of eigenfunctions belonging to the $\phi=1+\alpha$, i.e., the `odd' sector, we need to set $1+a-b=-n$, with $n=0,1,2,\dotsc$, so that $\Gamma(1+a-b)$ diverges, and consequently, the first term in \eqref{smallx.1} vanishes. Using the expression of $a$  and $b$ from \eref{eq:ab}, gives  the quantized eigenvalues corresponding to the odd eigenfunctions as
\begin{equation}
 \lambda_n^\mathrm{odd} = 2(\alpha+2)\mu n + (2\alpha+3)\mu \quad \text{where}\quad n=0,1,2,\ldots\, ,
\label{eq:odd-lambda}
\end{equation}
The associated odd eigenfunctions, denoted by $\chi_n(x)$ with $n=0,1, 2, \dotsc$,  are then obtained by setting $\lambda=\lambda_n^\mathrm{odd}$ and $a=b-1-n$ in \eref{eq:f-pot3}. Again, we can obtain the odd eigenfunctions for the whole space $x\in (-\infty, \infty)$ by making the eigenfunctions anti-symmetric. Using the identity $U(b-1-n,b,z)= z^{1-b} U(-n,2-b,z)$~\cite{abramowitz1965handbook} and then using the relation in~\eref{eq:UL}, we get
\begin{equation}
    \chi_n(x)=B_n\, x|x|^\alpha\, e^{-\frac{\mu |x|^{\alpha+2}} {\alpha+2}}\,  L_n^{(\frac{\alpha+1}{\alpha+2})}\left( \frac{2\mu\, |x|^{\alpha+2}}{\alpha+2}\right).
    \label{eq:chin}
\end{equation}
Normalizing the eigenfunctions over the whole space, i.e.,  $\int_{-\infty}^\infty\chi_m(x)\chi_n(x)\, dx=\delta_{m,n}$  and using the orthogonality relation~\eqref{eq:Lag-ortho}, we obtain
\begin{equation}
    B_n = \sqrt{\left(\frac{2\mu}{\alpha+2}\right)^{\gamma}\, \frac{\mu \, n!}{\Gamma(n+1+\gamma)}}\quad\text{with}\quad \gamma=\frac{\alpha+1}{\alpha+2}.
    \label{eq:Cn}
\end{equation}
As in the symmetric (even) case, if the eigenfunction is normalized over the half-space $x\in (0,\infty)$, then the normalization constant $B_n$ becomes $\sqrt{2}$ times the expression of $B_n$ in \eqref{eq:Cn}.

Using  $L_n^{(1/2)} (z)= (-1)^n \,H_{2n+1}(\sqrt{z})/(n! 2^{2n+1}\sqrt{z})$~\cite{abramowitz1965handbook}, it is easy to check that $\chi_n(x)$ in \eref{eq:chin} for $\alpha=0$, reduces to the odd eigenfunctions of a quantum harmonic oscillator,
given in \eref{eq:Hermite1} with $k=2n+1$.

\begin{figure}
    \centering
    \includegraphics[width=0.9\linewidth]{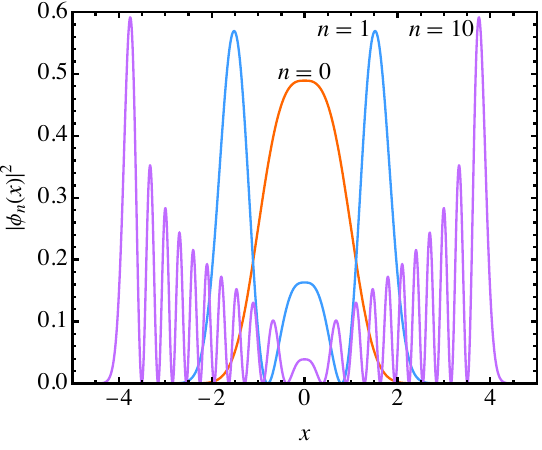}
    \includegraphics[width=0.9\linewidth]{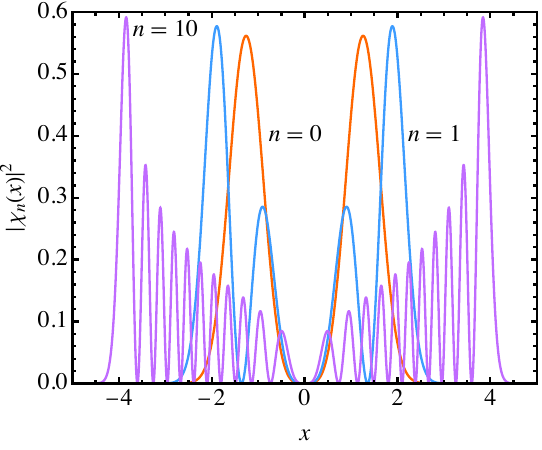}
    \caption{Plot of the square of the absolute values of the eigenfunctions $|\psi_n(x)|^2$, where $\psi_n(x)$ is given in \eref{eq:psi-even-odd}. The top panel is for the even eigenfunctions $\psi_{2n}(x)=\phi_n(x)$ in \eref{eq:phin} and the bottom panel is for the odd eigenfunctions $\psi_{2n+1}(x)=\chi_n(x)$ in \eref{eq:chin}. As we see from the figures, $|\psi_n(x)|^2$ is peaked away from the origin for all excited states, except for the ground state $|\psi_0(x)|^2 = |\phi_0(x)|^2$ (orange in top panel). We set $\alpha=\mu=1$. }
    \label{fig:eigenfunctions}
\end{figure}

\begin{figure}
    \centering
    \includegraphics[width=.9\linewidth]{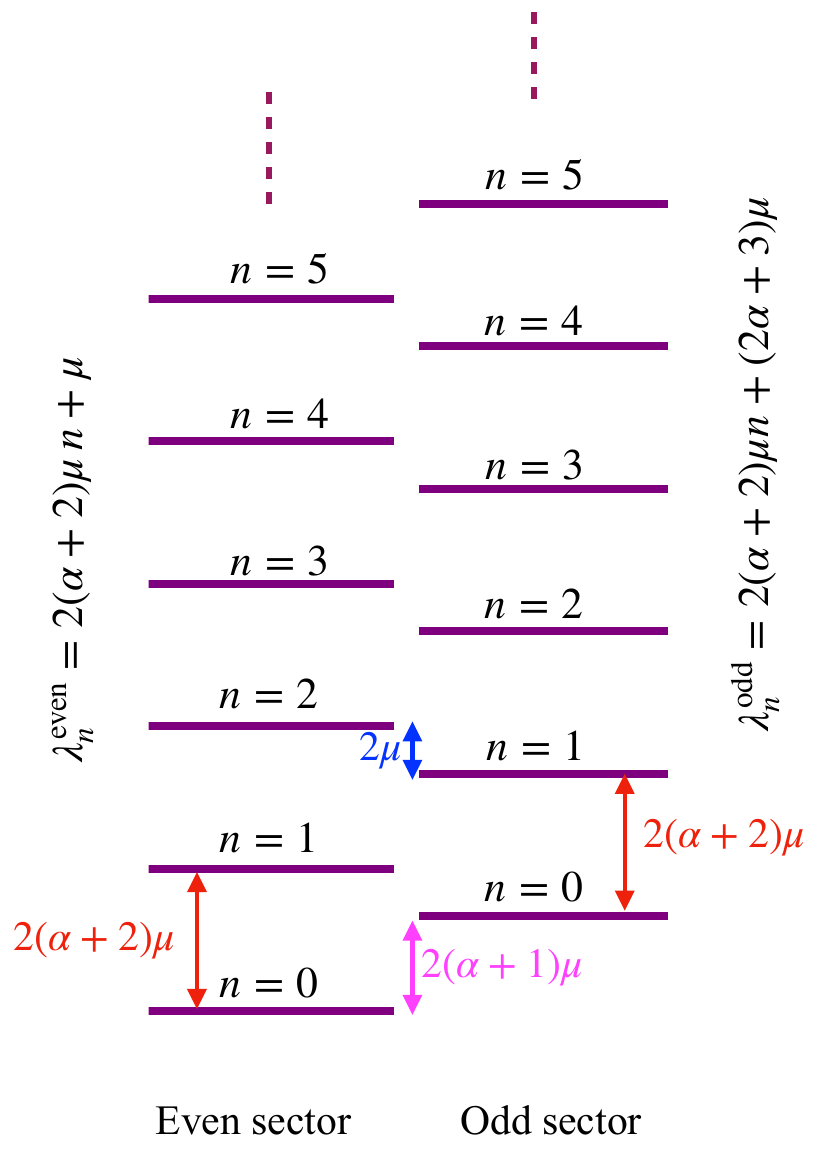}
    \caption{The eigenvalue spectrum $\lambda_k$ with $k=0, 1, 2, \dotsc$ is divided into even and odd sectors: $\lambda_{2n}=\lambda_n^\text{even}$ and $\lambda_{2n+1}=\lambda_n^\text{odd}$ with $n=0, 1, 2, \dotsc$, which are interlaced. The even and odd eigenvalues are given by Eqs.~\eqref{eq:even-lambda} and \eqref{eq:odd-lambda} respectively.}
    \label{fig:spectrum}
\end{figure}

To briefly summarize, the eigenfunctions $\psi_k(x)\equiv f_{E_k}(x)$ satisfying the Schr\"odinger equation~\eqref{eq:SE-timeInd-pot} with quantized energies $E_k = \lambda_k D_0$, where $k=0, 1, 2, \dotsc,\infty$ and the potential $V_\mathrm{ext}(x) = \mu^2 D_0 |x|^{\alpha+2}$, can be divided into even (symmetric) and odd (anti-symmetric) groups, such that 
\begin{equation}
   \psi_{2n}(x)=\phi_n(x)\quad\text{and}\quad \psi_{2n+1}(x)=\chi_n(x), 
   \label{eq:psi-even-odd}
\end{equation}
with $n=0,1,2,\dotsc,\infty$, where $\phi_n(x)$ and $\chi_n(x)$ are given by Eqs.~\eqref{eq:phin} and \eqref{eq:chin} respectively. \Fref{fig:eigenfunctions} shows the square of the absolute values of the eigenfucntion $|\psi_n(x)|^2$. We see from the figure that, for $\alpha>0$, the maximum value of $|\psi_n(x)|^2$ is away from the origin for all excited states, except for the ground state $|\psi_0(x)|^2 = |\phi_0(x)|^2$. 
The corresponding even and odd eigenvalues, 
\begin{equation}
\lambda_{2n}=\lambda_n^\text{even}\quad \text{and}\quad \lambda_{2n+1}=\lambda_n^\text{odd}\, ,
\label{eq:lambda-even-odd}
\end{equation}
with $n=0, 1, 2, \dotsc,\infty$, are given by  Eqs.~\eqref{eq:even-lambda}, and \eqref{eq:odd-lambda}, respectively. The gap between the successive odd and even eigenvalue pairs is given by $\lambda_n^\text{odd}-\lambda_n^\text{even} = 2 (\alpha+1)\mu$. On the other hand, the gap between the successive even and odd eigenvalue pairs is $\lambda_{n+1}^\text{even}-\lambda_n^\text{odd}=2\mu $. Therefore, the eigenvalues are equally spaced only at $\alpha=0$, which corresponds to the harmonic oscillator. \Fref{fig:spectrum} shows the interlacing of the even and odd spectra.

\subsection{Remarks on the corresponding classical stochastic problem}
\label{s:remarks}

It is useful to compare the quantum probabilities $|\psi_n(x)|^2$, given in \eref{eq:psi-even-odd} and illustrated in \fref{fig:eigenfunctions}, with the corresponding position distribution of a classical sluggish Brownian particle. Similar to the derivation of the sluggish Schr\"dinger equation from the microscopic hopping elements in \sref{s:derivation}, here, again, we start with a microscopic model of a random walk with inhomogeneous hopping rates. To this end, we consider a classical random walk on a lattice with lattice spacing $\delta$, an external potential $V_\text{ext}(x)$, and mobility $\mu(x)$.  In a small time interval $\Delta t$, the position $x(t)$ evolve by the following stochastic rules
\begin{equation}
    x(t) \to 
    \begin{cases}
        x(t)  +\delta & \text{with prob. }\, w_{+} \Delta t \\[3mm]
        x(t)  -\delta & \text{with prob. }\, w_{-}\Delta t  \\[3mm]
        x(t) & \text{with prob. }\,1 - (w_{+}+ w_{-})\Delta t 
        \label{eq:stochastic-rule}
    \end{cases}
\end{equation}
where $w_\pm\equiv w(x\to x\pm \delta)$ are position dependent rates for hopping from $x$ to $x\pm \delta$, which are given by
\begin{align}
    w (x\to y) & = J\bigl[(1-r) D(x) + r D(y) \bigr]  \notag \\ &+ \tilde{J} \bigl[(1-r) \mu(x) + r \mu(y)\bigr]\, \bigl[V_\text{ext}(x)-V_\text{ext}(y)\bigr].
    \label{eq:ratew}
\end{align}
Here, $0\le r \le 1$ is a parameter that dictates the discretization scheme. For example, $r=0$ and $r=1/2$ correspond to the It\^o and the Stratonovich rules of discretization, respectively. The coefficients $J$ and $\tilde{J}$ in \eref{eq:ratew} characterize the hopping strengths arising from two different mechanisms. 

Given the stochastic rules in \eref{eq:stochastic-rule}, in a small time interval $[t,t+\Delta t]$, the position distribution $P(x,t)$ evolves  according to 
\begin{align}
     P(x,t+\Delta t) &= P(x+\delta, t) \,w(x+\delta \to x)\,\Delta t \notag \\ 
     & + P(x-\delta, t)\, w(x-\delta \to x)\Delta t  \notag \\
     +  P(x,t) &\Bigl(1- \bigl[w(x\to x+\delta) + w(x\to x-\delta)\bigr]\Delta t\Bigr).
     \label{eq:masterE}
\end{align}
Now substituting the rates from \eref{eq:ratew} in \eref{eq:masterE}, expanding in Taylor series in $\delta$, and taking the continuum limit $\Delta t\to 0$,    $\delta\to 0$, $(J, \tilde{J})\to \infty$, keeping $J \delta^2=1$ and $2 \tilde{J}\delta^2=1$, we get
\begin{equation}
 \frac{\partial P(x,t)}{\partial t} = \frac{\partial}{\partial x} \left[D(x) \frac{\partial P (x,t)}{\partial x}    -F(x)  P(x,t) \right],
    \label{eq:FP0}    
\end{equation}
with the drift $F(x)$ given by
\begin{equation}
    F(x)=-(1-2r) D'(x) - \mu(x)\,V'_\text{ext}(x).
    \label{eq:F-r}
\end{equation}
The additional drift term $(1-2r) \,D'(x)$ in \eref{eq:F-r} disappears for the Stratonovich case $r=1/2$. This Stratonovich Fokker-Planck equation~\cite{vankampen2007stochastic, SL2003} 
\begin{equation}
    \frac{\partial P(x,t)}{\partial t} = \frac{\partial}{\partial x} \left[D(x) \frac{\partial P (x,t)}{\partial x}    + \mu(x)\, V_\mathrm{ext}'(x)\, P(x,t) \right],
    \label{eq:FP1}
\end{equation}
is precisely the classical analogue of the quantum sluggish Schr\"odinger equation~\eqref{eq:SE-slug-mod}.
It should be noted that while the operator $\partial_x D(x)\partial_x$ in \eref{eq:FP1} is self-adjoint like its quantum counterpart in \eref{eq:SE-slug-mod}, the operator involving the potential term is not self-adjoint, unlike the quantum case.  
If the potential $V_\mathrm{ext}(x)$ is confining, then the system eventually reaches a stationary state, $P(x,t\to\infty)\equiv P_\text{ss}(x)$, given by
\begin{equation}
    P_\text{ss}(x)\propto \exp\left( -\int^x \frac{\mu(y)}{D(y)}\, V_\text{ext}'(y)\, dy \right).
    \label{eq:Pss1}
\end{equation} 
If the mobility and the diffusion coefficient  satisfy the Einstein relation, $D(x)=\mu(x) k_\text{B} \, T$, where $k_\text{B}$ is the Boltzmann constant and $T$ is the temperature of the bath, the steady state is given by the Gibbs-Boltzmann distribution 
\begin{equation}
    P_\text{ss}(x) \propto \exp\left(-\frac{V_\text{ext}(x)}{k_\text{B}\, T}\right).
    \label{eq:Pss1b}
\end{equation}
On the other hand, if there is no Einstein relation between $\mu(x)$ and  $D(x)$, the stationary state depends on the functional forms of $\mu(x)$ and $D(x)$.
In particular, if we choose, $\mu(x)=1$, $D(x)=|x|^{-\alpha}$, and $V_\text{ext}(x)= |x|^{\alpha+2}$, then we get
\begin{equation}
    P_\text{ss} (x)\propto \exp\left(-\frac{(\alpha+2)}{2(\alpha+1)}\, |x|^{2(\alpha+1)}\right).
    \label{eq:Pss2}
\end{equation}
In both cases, i.e., Eqs.~\eqref{eq:Pss1b} and \eqref{eq:Pss2}, the stationary states have maxima at the origin. This is in contrast to the quantum case, where maxima of the quantum probabilities $|\psi_n(x)|^2$  for the excited states ($n>0$) are away from the origin, unless $\alpha=0$.

While $r\ne 1/2$ does not have a quantum analogue, it is nonetheless useful to analyze. In particular, for $r<1/2$, the additional drift term $- (1-2r)D'(x)$ pushes the particle away from the origin for $D(x)=|x|^{-\alpha}$. As we will show below for the It\^o case $r=0$, this outward push causes a depletion at the center.
For simplicity, we set the mobility $\mu(x)=1$. In that case, we recall the It\^o Fokker-Planck equation [$r=0$ in Eqs.~\eqref{eq:FP0}--\eqref{eq:F-r}] in the more standard form~\cite{SL2003, AZ2023}
\begin{equation}
  \frac{\partial P(x,t)}{\partial t} = \frac{\partial}{\partial x} \left[ \frac{\partial }{\partial x} D(x) P (x,t)   +  V_\mathrm{ext}'(x)\, P(x,t) \right],
    \label{eq:FP2}  
\end{equation}
However, we caution again that \eref{eq:FP2} is not the classical analogue of the sluggish Schr\"odinger equation~\eqref{eq:SE-slug-mod}, as the operator $\partial_x^2 D(x)$ in \eref{eq:FP2} is not self-adjoint, and therefore, there is no quantum equivalence. The stationary state of \eref{eq:FP2} equivalently, Eqs.~\eqref{eq:FP0}--\eqref{eq:F-r} with $r=0$] becomes
\begin{equation}
    P_\text{ss}(x)\propto \frac{1}{D(x)}\, 
    \exp\left( -\int^x \frac{V'(y)}{D(y)}\, \, dy \right).
    \label{eq:Pss3}
\end{equation}
If $D(x)=|x|^{-\alpha}$ and $V(x)= |x|^{\alpha+2}$, we get
\begin{equation}
   P_\text{ss} (x) \propto |x|^\alpha\,\exp\left(-\frac{(\alpha+2)}{2(\alpha+1)}\, |x|^{2(\alpha+1)}\right).
    \label{eq:Pss4} 
\end{equation}
 
 To conclude, for $D(x)=|x|^{-\alpha}$, since there is no outward drift away from the origin for $r \ge 1/2$, there is no depletion at the origin for the Stratonovich case ($r=1/2$). On the other hand, there is an outward drift away from the origin for the It\^o case ($r=0$). Therefore, it is surprising that our quantum sluggish model for $\alpha>0$, which is closer to the classical Stratonovich case, shows depletion near the origin for the quantum probabilities $|\psi_n(x)|^2$ in \fref{fig:eigenfunctions}, for the excited states ($n>0$).
Hence, our findings of this work have a purely quantum origin because the eigenfunctions for the
excited states are peaked away from the origin. We will see in \sref{s:density} that this depletion of the single-particle quantum probabilities near the origin results in a depletion in the density profile of a large number of  noninteracting fermions in the trap $V_\text{ext}(x)$.

\subsection{The complete wavefunction and the propagator}

The wavefunction $\psi(x,t)$ in \eref{eq:psixt-pot}, now becomes a discrete sum in terms of the eigenvalues and  eigenfunctions, 
\begin{align}
    \psi(x,t) = 
    \sum_{m=0}^\infty 
    C_m \, e^{-(iD_0/\hbar)\, \lambda_m t}\, \psi_m(x), 
    \label{eq:psi-pot4}
\end{align}
where $\psi_m(x)$ and $\lambda_m$ are given in Eqs.~\eqref{eq:psi-even-odd} and \eqref{eq:lambda-even-odd}, respectively. 

For a given initial condition $\psi(x,0)$, the coefficients $C_m$, with $m=0, 1, 2, \dotsc$, are given by
\begin{equation}
   \begin{split}
    C_{2n} &= \int_{-\infty}^\infty\, \psi_0^+ (x)\, \phi_n(x)\, dx,\\
    C_{2n+1} &= \int_{-\infty}^\infty\, \psi_0^- (x)\, \chi_n(x)\, dx,
\end{split} 
\label{eq:Cm}
\end{equation}
with $n=0, 1, 2, \dotsc$, where
\begin{equation}
    \psi_0^\pm(x) = \frac{\psi_0(x) \pm  \psi_0(-x)}{2}\, ,
\end{equation}
represents the symmetric and anti-symmetric components of the initial state $\psi_0(x)$, respectively.
Substituting $C_m$ from \eref{eq:Cm} in \eref{eq:psi-pot4}, we get
\begin{align}
    \psi(x,t)&=\int_{-\infty}^\infty\, dx_0\, \psi_0^+(x_0)\, G^+(x,t|x_0,0)\notag\\ &+ \int_{-\infty}^\infty\, dx_0\, \psi_0^-(x_0)\, G^-(x,t|x_0,0),
    \label{eq:time_evol}
\end{align}
where $G^\pm(x,t|x_0,0)$ are the symmetric and the anti-symmetric propagators, respectively, given by
\begin{align}
\label{eq:GS-pot}
 G^+(x,t|x_0,0)&= \sum_{n=0}^\infty \phi_n(x_0)\, \phi_n(x)\,   e^{-(iD_0/\hbar)\, \lambda_n^\mathrm{even}\, t},\\  
 \label{eq:GA-pot}
 G^-(x,t|x_0,0)&= \sum_{n=0}^\infty \chi_n(x_0)\, \chi_n(x)\,   e^{-(iD_0/\hbar)\, \lambda_n^\mathrm{odd}\, t}.
\end{align}
It turns out that the summations over $n$ in Eqs.~\eqref{eq:GS-pot} and \eqref{eq:GA-pot} can be carried out by using the Hardy--Hille formula~\cite{erdelyi1953higher}
\begin{align}
    \sum_{n=0}^\infty &\frac{n!}{\Gamma(n+1+\gamma)}\, L_n^{(\gamma)} \,(z_0) L_n^{(\gamma)}(z)\, u^n \notag \\
    &= \frac{1}{(z \,z_0 \,u)^{\gamma/2} (1-u)} e^{-\frac{(z+z_0) u}{1-u}}\, I_\gamma \left(\frac{2\sqrt{z \, z_0\, u}}{1-u}\right)\, ,
    \label{eq:HH}
\end{align}
where $I_\gamma(z)$ is the modified Bessel function of first kind of order $\gamma$~\cite{abramowitz1965handbook}. Using~\eref{eq:HH} and the forms of $\phi_n(x)$ and $\lambda_n^\mathrm{even}$ from Eqs.~\eqref{eq:phin} and \eqref{eq:even-lambda} respectively, we get the symmetric propagator as
\begin{equation}
    G^+(x,t|x_0,0)= \bar{G}^+\left(\frac{2\mu\, |x|^{\alpha+2}}{\alpha+2}, \frac{2\mu\, |x_0|^{\alpha+2}}{\alpha+2}, \frac{D_0\, \mu\,(\alpha+2)\, t}{\hbar}\right),
    \label{eq:GSs}
\end{equation}
where 
\begin{align}
    \bar{G}^+(z, z_0, \tau)  = \left(\frac{2\mu}{\alpha+2}\right)^{-\gamma}\mu \, \frac{ e^{\frac{i}{2}(z+z_0) \, \cot{\tau}}}{2 i \sin\tau}\,
    (z z_0)^{\gamma/2}\, I_{-\gamma}\left(\frac{\sqrt{zz_0}}{ i \sin\tau}\right),
    \label{eq:GSs2}
\end{align}
with $\gamma=(\alpha+1)/(\alpha+2)$.

Likewise, using~\eref{eq:HH} and the forms of $\chi_n(x)$ and $\lambda_n^\mathrm{odd}$ from Eqs.~\eqref{eq:chin} and \eqref{eq:odd-lambda} respectively, we get the anti-symmetric propagator as
\begin{align}
    G^-&(x,t|x_0,0) = \mathrm{sgn}(x)\, \mathrm{sgn}(x_0)\,\notag \\
    &\times \bar{G}^-\left(\frac{2\mu\, |x|^{\alpha+2}}{\alpha+2}, \frac{2\mu\, |x_0|^{\alpha+2}}{\alpha+2}, \frac{D_0\, \mu\,(\alpha+2)\, t}{\hbar}\right),
    \label{eq:GAs}
\end{align}
where
\begin{align}
    \bar{G}^- (z, z_0, \tau)  =\left(\frac{2\mu}{\alpha+2}\right)^{-\gamma}\,\mu \, \frac{ e^{\frac{i}{2}(z+z_0) \, \cot{\tau}}}{2 i \sin\tau}\,
    (z z_0)^{\gamma/2}\, I_{\gamma}\left(\frac{\sqrt{zz_0}}{ i \sin\tau}\right),
    \label{eq:GAs2}
\end{align}
with $\gamma=(\alpha+1)/(\alpha+2)$.

For the case $\alpha=0$, using $I_{\pm 1/2}(x) = (e^{x}\mp e^{-x})/\sqrt{2\pi x} $, it can be checked that $G^\pm (x,t|x_0,0)$ reduces to the corresponding symmetric and anti-symmetric propagators for the conventional quantum harmonic oscillator. In particular, $G^+ (x,t|x_0,0)+ G^-(x,t|x_0,0)$ gives the Mehler kernel~\cite{feynman2010quantum, KMS_2025}.

Having obtained the explicit forms of the propagators, namely, Eqs.~\eqref{eq:GSs} and \eqref{eq:GAs}, one can compute the time evolution of the wavefunction $\psi(x,t)$ using \eref{eq:time_evol}, for any initial state $\psi_0(x)$. Our next goal is to use the symmetric and the anti-symmetric eigenfunctions, $\phi_n(x)$ in \eref{eq:phin} and $\chi_n(x)$ in \eref{eq:chin}, respectively, and study the many-body ground state of $N$ noninteracting spinless sluggish fermions in an external trap $V_\mathrm{ext}(x)=\frac{1}{2}m_\mathrm{eff}\,\omega |x|^{\alpha+2}$. However, before that, we take a small digression to show that it is possible to reduce the eigenvalue equation in Eq.~\eqref{eq:f-diff-pot} to a standard Schr\"odinger equation with a constant effective mass $m_\text{eff}$ (as in the $\alpha=0$ case), albeit with a different effective potential.

\subsection{Mapping to a standard Schr\"odinger equation}
\label{s:mapping}

In this subsection, we show that it is possible to reduce the eigenvalue equation in Eq.~\eqref{eq:f-diff-pot} to a standard Schr\"odinger equation (as in the $\alpha=0$ case), but with a different effective potential $\tilde V_{\rm ext}(x)$, which turns out to be a sum of harmonic and inverse square potential. 

We start from Eq.~\eqref{eq:f-diff-pot} and restrict to $x>0$ as usual. Then we make the change of variables
\begin{equation}
    f_\lambda(x) = x^{\frac{\alpha}{4}}\tilde \psi_\lambda\left(\frac{\sqrt{2}}{\alpha+2}x^{\frac{\alpha+2}{2}}\right)\, .
    \label{eq:change_var1}
\end{equation}
Substituting this form into Eq.~\eqref{eq:f-diff-pot}, a straightforward computation leads to following eigenvalues equation for $\tilde\psi_\lambda(y)$
\begin{equation}
    -\tilde \psi''(y) + \tilde V(y) \tilde \psi_\lambda(y) = \lambda \tilde \psi_\lambda(y)\,,
    \label{eq:psi_til_eq}
\end{equation}
where 
\begin{equation}
    \tilde V(y) = \frac{1}{2}\mu^2 (\alpha+2)^2 y^2 +\frac{\alpha(3\alpha + 4)}{8(\alpha+2)^2}\frac{1}{y^2}\,.
    \label{eq:Veff2}
\end{equation}
To solve this eigenvalue problem uniquely in the region $y>0$, one has to specify the boundary conditions as $y\to\infty$ and as $y\to0$. The boundary condition as $y\to\infty$ is standard, namely, the eigenfunction $\tilde \psi_\lambda(y)$ must vanish, since $\tilde V(y)$ is a confining potential. The boundary condition as $y\to0$ is trickier. To find the appropriate boundary condition as $y\to0$, we recall the mapping in Eq.~\eqref{eq:change_var1} and the boundary conditions satisfied by the even and odd sectors by $f_\lambda(x)$ as $x\to 0$ in the original $x$-coordinate, as in Eq.~\eqref{smallx.1}. Taking the $x\to 0$ limit in Eq.~\eqref{eq:change_var1} and using Eq.~\eqref{smallx.1}, it follows that even and odd sectors for $f_\lambda(x)$ correspond respectively to the following behavior for $\tilde \psi_\lambda(y)$ as $y\to 0$
\begin{equation}
    \tilde \psi_\lambda(y)\sim \begin{cases}
        y^{-\frac{\alpha}{2(\alpha+2)}} & \text{(even)},
        \\[3mm]
        y^{\frac{3\alpha+4}{2(\alpha+2)}} & \text{(odd)}.
    \end{cases}\,\label{eq:psi_b0}
\end{equation}
Thus, the eigenfunction $\tilde\psi_\lambda(y)$ vanishes as $y\to 0$ for the odd sector, while it diverges as $y\to 0$ for the even sector. 

In fact, the eigenvalue problem in Eq.~\eqref{eq:psi_til_eq} with the harmonic plus inverse square potential in Eq.~\eqref{eq:Veff2} was studied in Ref.~\cite{nadal2009}, in the context of fluctuating interfaces in the presence of a hard wall. However, in that problem, due to the hard wall boundary condition at $y=0$, only the Dirichlet boundary condition, $\tilde \psi(y=0)=0$, i.e., the second line in Eq.~\eqref{eq:psi_b0}, is present. Due to this, in that problem, the even sector was absent, and the spectrum consisted only of the odd sector. However, in our problem, which lives in the $x$-space, both sectors are present under the mapping to the $y$ coordinates. Hence, in our problem, the full spectrum consists of both the even and the odd eigenfunctions, unlike in the fluctuating interface problem. Later, we will see a consequence of this fact when we consider $N$ non-interacting sluggish fermions in an external potential $V_{\rm ext}(x)$ as given in Eq.~\eqref{eq:V-eff}. 

In this respect, let us make a brief remark on $N$ non-interacting fermions that will be explained later in detail in Section \ref{s:mbgs}. The quantum joint probability density function of the positions of $N$ non-interacting fermions in the ground state of the potential $\tilde V_{\rm ext}(y)$ in Eq.~\eqref{eq:Veff2} (with a hard wall at $y=0$) can be exactly mapped to the joint distribution of $N$ eigenvalues of Wishart-Laguerre random matrix \cite{nadal2009}. Furthermore, in the large $N$ limit, the scaled kernel for the fermions near the edge at $y=0$ is given by the Bessel kernel \cite{forrester2010log, Lacroix2018}. This kernel does correspond to only the odd sector of the spectrum of the sluggish quantum problem in $x$-coordinates in Eq.~\eqref{eq:f-diff-pot}. In Section \ref{s:mbgs}, we will see that for $N$ non-interacting sluggish fermions, one needs to retain both sectors. This will lead to a new kernel, which is a sum of two Bessel kernels, corresponding to even and odd sectors, respectively.

Let us now return to the original $x$-space, and 
 study the many-body ground state of $N$ noninteracting spinless sluggish fermions in an external trap $V_\mathrm{ext}(x)=\frac{1}{2}m_\mathrm{eff}\,\omega |x|^{\alpha+2}$.
However, before studying the $\alpha>0$ case, it is useful to remind readers of the $\alpha=0$ case, namely, noninteracting spinless fermions in a harmonic trap, which has been widely studied and subject of growing interest. Therefore, in the next section, we recapitulate the ground-state properties of $N$ noninteracting spinless fermions.

\section{Recapitulation of the ground state properties of $N$ noninteracting spinless fermions in a harmonic potential with $\alpha=0$.}
\label{s:recap}

In this section, we recapitulate the properties of the many-body ground state of a system of $N$ noninteracting spinless fermions in a harmonic potential~\cite{Dean2019review}. We consider the $\alpha=0$ case, i.e., $D(x) = D_0= \hbar^2/(2m_{\mathrm{eff}})$ and 
$V_\mathrm{ext}(x)=\frac{1}{2}m_\mathrm{eff}\omega^2 \,x^2$. Let us consider $N$ spinless noninteracting fermions in the one-dimensional trapping potential $V_\mathrm{ext}(x)$. The system is described by the $N$-body Hamiltonian $H_N = \sum_{j=1}^N  h_j$ where $ h_j =  h( x_j,  p_j)$ is a single-particle Hamiltonian of the form
\begin{equation}
\label{eq:H1d}
 h = \frac{ p^2}{2\,m_\mathrm{eff}} + V_\mathrm{ext}(x) \;.
\end{equation}
Let $\psi_k(x)$ be the $k$-th single-particle eigenfunction ($k = 0,1,2, \dotsc$) with eigenvalue $\epsilon_k$, i.e.,
\begin{equation} 
\label{eq:ef}
 h \, \psi_k(x) = \epsilon_k \psi_k(x) \;.
\end{equation} 
The ground state of an $N$-particle system is obtained by occupying the lowest $N$ single-particle energy levels, placing one fermion in each level, in accordance with the Pauli exclusion principle. As a result, the many-body ground-state wavefunction takes the form of a Slater determinant, given by
\begin{equation}
\label{eq:Slater}
\Psi_0(x_1, \cdots, x_N) = \frac{1}{\sqrt{N!}} \, \det_{1\leq \,j,\,l\, \leq N}\psi_{l-1}(x_j)  \;,
\end{equation}
with the associated energy $E_0 = \sum_{k=0}^{N-1} \epsilon_k$. From~\eref{eq:Slater}, the quantum joint probability density function (JPDF) is then given by
\begin{equation}
\label{eq:jpdf}
P(x_1, \cdots, x_N) = |\Psi_0(x_1, \cdots, x_N)|^2 = \frac{1}{N!} \left|\det_{1\leq \,j,\,l\, \leq N} \psi_{l-1}(x_j) \right|^2 \;.
\end{equation}
The single-particle eigenfunctions $\psi_k(x)$ are given by
 \begin{equation}
 \label{eq:Hermite}
\psi_k(x) = \left[ \frac{\sqrt{\mu}}{\sqrt{\pi} 2^k k!}\right]^{1/2} \, e^{-\frac{\mu\,x^2}{2}} H_k(\sqrt{\mu} \,x) \;,
\end{equation}
 where $ \mu=m_\mathrm{eff}\,\omega/\hbar$  and  $H_k(z)$ is the  Hermite polynomial of degree $k$. The associated single-particle energy levels are given by $\epsilon_k = (k+1/2)\hbar \omega$. To construct the Slater determinant, we take the first $N$ energy levels labeled by $k=0,\cdots, N-1$. In the Slater determinant, the Gaussian factors come out of the determinant, leaving us to compute the determinant of a matrix consisting of Hermite polynomials. The Hermite polynomials $H_0(z), H_1(z), \cdots, H_{N-1}(z)$ provide a basis for polynomials of degree $N-1$, and by manipulating the rows and columns, the determinant can be reduced to a Vandermonde determinant. Hence, we can evaluate the Slater determinant explicitly to obtain
\begin{equation}
\label{eq:JPDF}
P(x_1, \cdots, x_N) = \frac{1}{z^{\rm GUE}_N} e^{-\mu \sum_{i=1}^N x_i^2}\,\prod_{i<j}(x_i-x_j)^2 \;.
 \end{equation}
Here, $z^{\rm GUE}_N$ is a normalization constant and the subscript ``GUE'' denotes the resemblance to the Gaussian Unitary Ensemble. In fact, \eref{eq:JPDF} is the joint distribution of the eigenvalues of a $N \times N$ GUE matrix of RMT~\cite{mehta2004random, forrester2010log} up to the rescaling factor of $\sqrt{\mu}$.  Clearly, the Vandermonde square term $\prod_{i<j}(x_i-x_j)^2$ \eref{eq:JPDF} provides an effective repulsion between any pair of fermions coming purely from the Pauli exclusion principle. Thus, even though the fermions are noninteracting to start with, their quantum statistics provides an effective pairwise repulsion.

We next discuss the kernel, its scaling behavior, and the density profile for the $\alpha=0$ case. This problem has the so-called determinantal structure. For noninteracting fermions in a harmonic potential, all the information about the quantum spatial fluctuations is contained in the JPDF in~\eref{eq:JPDF}. Of common interest are the $n$-point spatial correlation functions denoted by $R_n(x_1, \cdots, x_n)$, with $1 \leq n \leq N$, which are given by the different marginals of the full JPDF, i.e.,
\begin{eqnarray}
\label{eq:corr} 
R_n(x_1, \cdots, x_n) &=&  \frac{N!}{(N-n)!} \int d{x}_{n+1} \cdots \int d {x}_N \nonumber  \\ && \times P(x_1, \cdots, x_n, x_{n+1}, \cdots, x_N) \;,
\end{eqnarray}
where the integrals over the positions $x_i$'s run over their full domain. The particular case of  $n=1$ gives
\begin{equation} 
\label{eq:dens}
R_1(x) = N \, \int dx_2 \cdots \int d x_N P(x, x_2, \cdots, x_N) \;,
\end{equation}
which is directly related to the average density of fermions in the ground state 
\begin{equation}
     \rho_N(x) = \frac{1}{N}\left\langle \sum_{i=1}^N \delta(x-x_i) \right\rangle_{\mathrm GS}  \;,
     \label{eq:master-density}
\end{equation}
via
\begin{equation} 
\label{eq:R1_dens}
R_1(x) = N\,\rho_N(x) \;, 
\end{equation}
where $\langle \cdots \rangle_{\mathrm GS}$ denotes an average in the ground state. The density $\rho_N(x)$ in~\eref{eq:master-density} is normalized to unity such that $\rho_N(x) dx$ denotes the
average fraction of fermions in the region $[x, x+dx]$, in the ground state. The JPDF can be written in terms of the kernel as
\begin{equation}
\label{eq:jpdf_kernel}
P(x_1, \cdots, x_N) = \frac{1}{N!} \det_{1\leq \,j,\,l\, \leq N} K_N(x_j, x_l) \;,
\end{equation}
where the kernel $K_N(x,y)$ is defined by
\begin{equation}
\label{eq:kernel}
    K_N(x,y) = \sum_{k=0}^{N-1}  \psi^*_k(x) \psi_k(y) \;,
\end{equation}
The $n$-point correlation function $R_n(x_1, \cdots, x_n)$ can be written as an $n \times n$ determinant \cite{mehta2004random,forrester2010log} \begin{equation}
\label{eq:det_struc}
R_n(x_1, \cdots, x_n) = \det_{1 \leq \, j,\, l \, \leq n} K_N(x_j,x_l) \;,
\end{equation}
In fact, this is a special property of noninteracting spinless fermions
in a confining potential in its ground state. Such point processes $\{x_1,x_2,\ldots, x_N\}$ where all
correlation functions  can be expressed as determinants of a single quantity involving
only a pair of them, namely
the kernel $K_N(x_i,x_j)$ as in \eref{eq:det_struc},  are known as determinantal point processes (DPP)~\cite{forrester2010log}. Consequently, for a DPP, one just needs to determine the kernel, and all other observables can, in principle, be expressed in terms of this kernel.

From~\eref{eq:det_struc} and \eref{eq:R1_dens}, it is easy to see that 
\begin{equation}
\label{eq:kxx}
\rho_N(x) = \frac{1}{N} K_N(x,x) = \frac{1}{N} \sum_{k=0}^{N-1} |\psi_k(x)|^2 \;.
\end{equation}
For the harmonic potential $V(x) = \frac{1}{2}m_\mathrm{eff}\,\omega^2 x^2$, the limiting density profile is the well-known Wigner semi-circle, which takes the scaling form~\cite{mehta2004random, forrester2010log}
\begin{equation}\label{wigner}
\rho_N(x) \approx 
\frac{\sqrt\mu}{\sqrt{N}} 
f_W\left(\frac{\sqrt\mu\, x}{\sqrt{N}} \right) ~~ \text{with}~~ 
f_W(z) = \frac{1}{\pi}\sqrt{2-z^2} \;,
\end{equation}
where recall the $ \mu=m_\mathrm{eff}\,\omega/\hbar$

The kernel in the bulk has the scaling form~\cite{forrester2010log, Dean2019review}
\begin{equation}
    K_N(x+x', x+y') = \frac{1}{\ell(x)}\, K_\mathrm{bulk} \left(\frac{|x'-y'|}{\ell(x)}\right), 
\end{equation}
where 
\begin{equation}
    \ell(x)= \frac{1}{N \rho_N(x)}\quad\text{and}\quad K_\mathrm{bulk}(w) = \frac{\sin \pi w}{\pi w}\, .
    \label{eq:SineK}
\end{equation}
On the other hand, near the edge, say the right edge, the kernel has the scaling form~\cite{PJF93, forrester2010log, Dean2019review}
\begin{equation}
    K_N(x,y)= \frac{1}{W_N} \, K_\text{edge} \left(\frac{x-x_\text{edge}}{W_N}, \frac{y-x_\text{edge}}{W_N}\right).
    \label{eq:KNAiry0}
\end{equation}
where $W_N=N^{-1/6}/\sqrt{2\mu}$, $x_\text{edge}=\sqrt{2N/\mu}$, and
\begin{equation}
    K_\text{edge} (u,v)= \frac{\text{Ai}(u)\text{Ai}'(v) - \text{Ai}'(u)\text{Ai}(v)}{u-v}\, 
    \label{eq:AiryK0}
\end{equation}
The kernels $K_\text{bulk}(w)$ and $K_\text{edge}(u,v)$, in Eqs.~\eqref{eq:SineK} and \eqref{eq:AiryK0} are respectively called `sine' and the
`Airy' kernel~\cite{forrester2010log}.

Having recapitulated the $\alpha=0$ case, in the next section (\sref{s:mbgs}), we discuss the problem of $N$ noninteracting sluggish fermions ($\alpha>0$) in the external potential  $V_\mathrm{ext} (x)= \frac{1}{2} m_\mathrm{eff}\omega^2\, |x|^{\alpha+2}$.

\section{The many-body ground state of $N$ noninteracting spinless sluggish fermions in an external trap $V_\mathrm{ext}(x)=\frac{1}{2} m_\mathrm{eff}\,\omega^2 |x|^{\alpha+2}$}
\label{s:mbgs}

As recapitulated in \sref{s:recap}, due to the Pauli exclusion principle, even noninteracting fermions in external confining potentials often lead to highly non-trivial strong correlations~\cite{Dean2019review}. In particular, the JPDF of the positions of harmonically trapped noninteracting fermions in the ground state is deeply connected to the JPDF of the eigenvalues of Gaussian unitary random matrices. This connection has led to a plethora of activities.
Therefore, it is natural to explore the properties of $N$ noninteracting sluggish fermions in an external trap. As we discussed in \sref{s:sluggish-potential}, the natural extension of the harmonic potential to the sluggish particle is $V_\mathrm{ext}(x)=\frac{1}{2}m_\mathrm{eff}\,\omega^2 |x|^{\alpha+2}$. We remark that the many-body ground state of $N$ noninteracting bosons is trivial, and is given by $\Psi_0(x_1, x_2, \dotsc, x_N)=\prod_{i=1}^N \psi_0(x_i)=\prod_{i=1}^N \phi_0(x_i)$, where $\phi_0(x)$ is given in \eref{eq:phin} with $n=0$. In the case 
of bosons in the ground state, the quantum probability distribution $|\psi_0(x_1,x_2,\ldots, 
x_N)|^2$ thus factorizes, making the positions independent and identically distributed (IID) 
random variables. In this case, the observables in the ground state can be computed exactly 
using the standard techniques of IID variables~\cite{SM_Book} and we do not discuss this further here.

\subsection{Joint probability density function of the positions and the quantum correlation kernel}
\label{s:JPDF-kernel}

We consider putting $N$ noninteracting sluggish fermions in the potential $V_\mathrm{ext}(x)= \frac{1}{2}m_\mathrm{eff}\,\omega^2 |x|^{\alpha+2}$ at zero temperature. The many-body ground state wavefunction can be constructed using the Slater determinant as
\begin{equation}
    \Psi_0(x_1,x_2, \dotsc, x_N)= \frac{1}{\sqrt{N!}} \det[\psi_{i-1}(x_j)]_{ 1\le i,j\le N} \, .
    \label{eq:MB-psi}
\end{equation}
where $\psi_n(x)$ denotes the $n$-th eigenfunction of a single sluggish particle in the potential $V_\mathrm{ext}(x)$ [see \eref{eq:psi-even-odd}]. Consequently, the many-body JPDF of the $N$ noninteracting fermions in the ground state is given by
\begin{equation}
    P(x_1, x_2, \dotsc, x_N) = |\psi_0(x_1, x_2, \dotsc, x_N)|^2 .
    \label{eq:JPDF1}
\end{equation}
Using \eref{eq:MB-psi} in \eref{eq:JPDF1}, and the multiplication properties of determinants, $\det(A)\det(B)=\det(AB)$,we get 
\begin{equation}
  P(x_1, x_2, \dotsc, x_N)  = \frac{1}{N!}\, \det[K_N(x_i,x_j)]_{1\le i,j \le N}, 
  \label{eq:JPDF-MB}
\end{equation}
with  
\begin{equation}
K_N(x,y)=\sum_{k=0}^{N-1} \psi_k(x) \psi_k(y) .
    \label{eq:kernel-MB}
\end{equation}
The function $K_N(x_i,x_j)$ in \eref{eq:kernel} is known as the 
quantum correlation kernel, which is 
one of the most central objects in many-body quantum mechanics, especially for noninteracting fermions at zero temperature.

Using the even and odd eigenfunctions [see \eref{eq:psi-even-odd}], $\phi_n(x)$ and $\chi_n(x)$, given in Eqs.~\eqref{eq:phin} and \eqref{eq:chin} respectively, the correlation kernel in \eref{eq:kernel-MB} becomes
\begin{equation}
    K_N(x,y) = K_{\lceil\frac{N}{2}\rceil}^+(x,y) + K_{\lfloor\frac{N}{2}\rfloor}^-(x,y),
     \label{eq:kernel-MB2}
\end{equation}
where $\lceil x\rceil$ and $\lfloor x\rfloor$ are  ceiling and floor functions respectively, and 
\begin{align}
\label{eq:K+}
K_M^+(x,y)&=\sum_{n=0}^{M-1} \phi_n(x)\phi_n(y)\\
\label{eq:K-}
K_M^-(x,y)&=  \sum_{n=0}^{M-1}\chi_n(x)\chi_n(y)\, .
\end{align}

To perform the summations in Eqs.~\eqref{eq:K+} and \eqref{eq:K-}, we recall the Christoffel–Darboux formula~\cite{abramowitz1965handbook} for sequence of orthogonal polynomials $\{f_n\}$, 
\begin{equation}
    \sum_{n=0}^m \frac{f_n(x) f_n(y)}{h_n}= \frac{k_m}{h_m k_{m+1}}\, \frac{f_n(y)f_{n+1}(x) - f_{n+1}(y) f_n(x)}{x-y},
    \label{eq:CDformula}
\end{equation}
where $h_n$ are squared norms of the polynomials and $k_m$'s are the lead coefficients, i.e., the coefficient of $x^m$ in a polynomial of degree $m$ (coefficient of the leading term). In our case, the involved orthogonal polynomials are generalized Laguerre polynomials $L_n^{(\pm \gamma}(z)$, for which 
$k_n$ and $h_n$ are given by [see Eqs.~\eqref{eq:Laguerre} and \eqref{eq:Lag-ortho}]
\begin{equation}
    k_n= \frac{(-1)^n}{n!}\quad\text{and}\quad h_n=\frac{\Gamma(n+1\pm \gamma)}{n!}\, .
\end{equation}

Therefore, substituting $\phi_n(x)$ and $\chi_n(x)$ from Eqs.~\eqref{eq:phin} and \eqref{eq:chin} in \eqref{eq:K+} and \eqref{eq:K-} respectively, and using the Christoffel–Darboux formula~\eqref{eq:CDformula}, we get
\begin{align}
\label{eq:K+s}
K_M^+(x,y) & = \bar{K}_M^+\left(\frac{2\mu\, |x|^{\alpha+2}}{\alpha+2}, \frac{2\mu\, |y|^{\alpha+2}}{\alpha+2}\right),   \\
\label{eq:K-s}
K_M^-(x,y) & = \mathrm{sgn}(x)\, \mathrm{sgn}(y)\,\bar{K}_M^-\left(\frac{2\mu\, |x|^{\alpha+2}}{\alpha+2}, \frac{2\mu\, |y|^{\alpha+2}}{\alpha+2}\right),
\end{align}
where 
\begin{align}
    &\bar{K}^+_M(z_1,z_2) = 
    \left(\frac{2\mu}{\alpha+2}\right)^{-\gamma}\,\mu \,   
     \frac{\Gamma(M+1)}{\Gamma(M-\gamma)} \,e^{-\frac{1}{2}(z_1+z_2)}\,\notag\\ &\quad\quad\quad\times \frac{L_{M-1}^{(-\gamma)} (z_1)\, L_{M}^{(-\gamma)} (z_2)- L_{M}^{(-\gamma)} (z_1)\, L_{M-1}^{(-\gamma)} (z_2)}{z_1-z_2},
      \label{eq:KN1}
      \intertext{and}
&\bar{K}^-_M(z_1,z_2) = 
    \left(\frac{2\mu}{\alpha+2}\right)^{-\gamma}\,\mu \,   
     \frac{\Gamma(M+1)}{\Gamma(M + \gamma)} \, (z_1 z_2)^\gamma \,e^{-\frac{1}{2}(z_1+z_2)}\, \notag\\
     &\quad\quad\quad\times \frac{L_{M-1}^{(\gamma)} (z_1)\, L_{M}^{(\gamma)} (z_2)- L_{M}^{(\gamma)} (z_1)\, L_{M-1}^{(\gamma)} (z_2)}{z_1-z_2}  . 
     \label{eq:KN2}
\end{align}
Thus, we have obtained the exact form of the correlation kernel explicitly for any finite $N$. Using these expressions~\eqref{eq:KN1} and \eqref{eq:KN2}, we can, in principle, obtain the JPDF from \eref{eq:JPDF-MB}. However, in practice, it is hard to evaluate the determinant \eref{eq:JPDF-MB}, for our case, to obtain an explicit form of the JPDF. Interestingly, an explicit form of the JPDF can be obtained in the half-space, which we relegate to \sref{s:JPDF-half}. The density profile can be obtained directly from the correlation kernel, which we discuss next.

\subsection{Density profile}
\label{s:density}

\begin{figure}
    \includegraphics[width=0.9\linewidth]{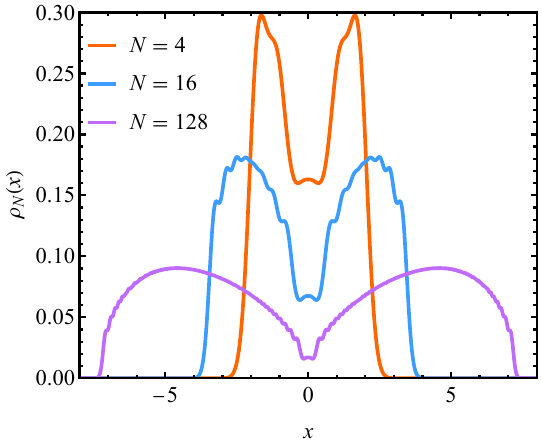}
    \caption{Density profile $\rho_N(x)$ [see~\eref{eq:density}], obtained by adding Eqs.~\eqref{eq:KN2a1} and \eqref{eq:KN2a2} for different numbers of fermions~($N$) with $\mu=1$ and $\alpha=1$.}
    \label{fig:fig1}
\end{figure}

Since the fermions obey the Pauli exclusion principle, the spatial density profile of noninteracting fermions can be quite non-trivial. This density (normalized to unity) can be obtained from the correlation kernel as [see \eref{eq:kxx}]~\cite{mehta2004random, forrester2010log, Dean2019review}
\begin{equation}
   \rho_N(x)=  \frac{1}{N}\lim_{y\to x}\, K_N(x,y) = \rho_N^+(x) + \rho_N^-(x)
    \label{eq:density}
\end{equation}
with 
\begin{equation}
    \rho_N^+(x) = \frac{1}{N} \lim_{y\to x}  K_{\lceil\frac{N}{2}\rceil}^+(x,y) ~\text{and}~ \rho_N^-(x)= \frac{1}{N} \lim_{y\to x}  K_{\lfloor\frac{N}{2}\rfloor}^-(x,y)\, ,
    \label{eq:rho_pm}
\end{equation}
where
$K_M^\pm (x,y)$  defined in Eqs.~\eqref{eq:K+} and \eqref{eq:K-} respectively. Using the expressions of $K_M^\pm(x,y)$ from Eqs.~\eqref{eq:K+s}--\eqref{eq:KN2}
and the derivatives of generalized Laguerre polynomials~\cite{gradshteyn2014table}
\begin{equation}
    \frac{d}{dz} L_M^{(\gamma)} (z) = - L_{M-1}^{(\gamma+1)}(z), 
    \label{eq:dLaguerre}
\end{equation}
we obtain $\rho_N^\pm(x)$ as
\begin{align}
\label{eq:KN2a1}
    &\rho_N^+(x) = \frac{1}{N}\left(\frac{2\mu}{\alpha+2}\right)^{-\gamma}\,\mu \,   e^{-z}  \, \frac{\Gamma(\lceil\frac{N}{2}\rceil+1)}{\Gamma(\lceil\frac{N}{2}\rceil-\gamma)} \,\notag\\
    &\quad\times\,\left[L_{\lceil\frac{N}{2}\rceil-1}^{(-\gamma+1)} (z)\, L_{\lceil\frac{N}{2}\rceil-1}^{(-\gamma)} (z)- L_{\lceil\frac{N}{2}\rceil-2}^{(-\gamma+1)} (z)\, L_{\lceil\frac{N}{2}\rceil}^{(-\gamma)} (z)\right],
    \intertext{and}
   &\rho_N^-(x) = 
   \frac{1}{N}\left(\frac{2\mu}{\alpha+2}\right)^{-\gamma}\,\mu \,   z^{2\gamma}\, e^{-z} \,
    \frac{\Gamma(\lfloor\frac{N}{2}\rfloor+1)}{\Gamma(\lfloor\frac{N}{2}\rfloor+\gamma)}\notag\\
&\quad\times\,\left[L_{\lfloor\frac{N}{2}\rfloor-1}^{(\gamma+1)} (z)\, L_{\lfloor\frac{N}{2}\rfloor-1}^{(\gamma)} (z)- L_{\lfloor\frac{N}{2}\rfloor-2}^{(\gamma+1)} (z)\, L_{\lfloor\frac{N}{2}\rfloor}^{(\gamma)} (z)\right]\, ,
 \label{eq:KN2a2}
\\[5mm]
    &\text{with}\quad z=\frac{2\mu\, |x|^{\alpha+2}}{\alpha+2}.
\end{align}
Equations~\eqref{eq:KN2a1} and \eqref{eq:KN2a2}, along with \eref{eq:density}, give the exact density profile of the fermionic gas for any  $N$, in the full space  $x\in (-\infty, \infty)$.  In \fref{fig:fig1}, we plot the density profile $\rho_N(x)$ by adding Eqs.~\eqref{eq:KN2a1} and \eqref{eq:KN2a2}. We notice a suppression of fermionic density close to the origin. The nonzero contribution to the density at the origin stems from the even sector, i.e., $\rho_N^+(x)$ in \eref{eq:KN2a1}, and is given by
\begin{equation}
    \rho_N^+(0) = \frac{2^{\frac{1}{\alpha +2}-2} \left(\frac{\mu }{\alpha +2}\right)^{\frac{1}{\alpha +2}} \Gamma \left(\lceil\frac{N}{2}\rceil+\frac{1}{\alpha +2}\right)}{\Gamma^2 \left(1+\frac{1}{\alpha +2}\right) \Gamma \left(\lceil\frac{N}{2}\rceil+1\right)}.
    \label{eq:density0}
\end{equation}
For large $N$, using Stirling's approximations in \eref{eq:density0}, we find that the density at the origin decays as $\rho_N^+(0) \sim N^{-\gamma}$ with $\gamma=(\alpha+1)/(\alpha+2)$, implying that the density must vanish at the origin in the large-$N$ scaling limit. As discussed in \sref{s:remarks}, this depletion of density near the origin (see \fref{fig:fig1}) is of purely quantum origin, because the eigenfunctions of the excited states are peaked away from the origin, as shown in \fref{fig:eigenfunctions}.

The large-$N$ limit of the density profile can be taken in two ways. One way is by taking the large $N$ limit directly in Eqs.~\eqref{eq:KN2a1} and \eqref{eq:KN2a2}. In this section, we obtain the large $N$ limit of the $\rho_N(x)$ in another way, by using the expressions of $K_M^\pm (x,y)$, given in Eqs.~\eqref{eq:K+} and \eqref{eq:K-}, in \eqref{eq:rho_pm}, i.e., 
\begin{equation}
   \rho_N^+(x) =  \frac{1}{N}\sum_{n=0}^{\lceil\frac{N}{2}\rceil-1} \phi_n^2(x),\quad  
   \rho_N^-(x) =  \frac{1}{N} \sum_{n=0}^{\lfloor\frac{N}{2}\rfloor-1}\chi_n^2(x)\, .
    \label{eq:density2}
\end{equation}
Since the energy $E_n$ scales linearly with the level $n$ [see Eqs.~\eqref{eq:even-lambda} and \eqref{eq:odd-lambda}], the total energy of a system with $N$ noninteracting fermions $\sum_{n=0}^{N-1} E_n\sim N^2$. On the other hand, the energy associated with the external trap $\sum_{i=1}^N |x_i|^{\alpha+2} \sim N |x|^{\alpha+2}$, where $x$ denotes typical positions. Since these two energies are of the same order, the typical size of the system $x$ must scale as $N^\frac{1}{\alpha+2}$. Therefore, in the large $N$ limit, it is natural to anticipate the following scaling of the density profile
\begin{align}
    \rho_N(x) = \frac{1}{N^\frac{1}{\alpha+2}}\,\bar{\rho} \left(\frac{x}{N^\frac{1}{\alpha+2}}\right), \quad 
    \rho_N^\pm(x) = \frac{1}{N^\frac{1}{\alpha+2}}\,\bar{\rho}^\pm  \left(\frac{x}{N^\frac{1}{\alpha+2}}\right).
    \label{eq:rho-scaling}
\end{align}

\begin{figure}
\includegraphics[width=0.9\linewidth]{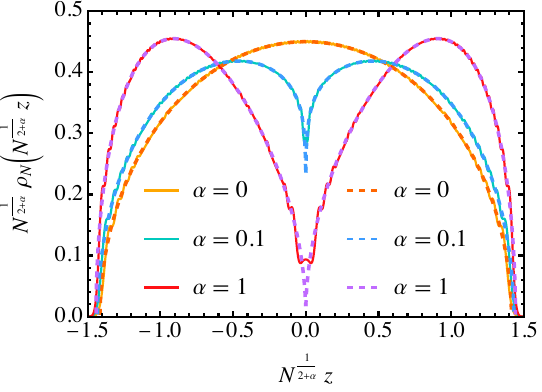}
    \caption{The solid lines plot the scaled density profile obtained from Eqs.~\eqref{eq:density}, \eqref{eq:KN2a1} and \eqref{eq:KN2a2},  for $N=100$, $\mu=1$, and different values of $\alpha$. The dashed lines plot scaling functions $\bar{\rho}(z)$ given in \eref{eq:rho-scal} with corresponding $\alpha$, for comparison. Note that the $\alpha=0$ case corresponds to the well-known Wigner semi-circle, given in \eref{eq:Wigner}. }
    \label{fig:fig2}
\end{figure}

Making a change of variable $n=Nu/2$, for large $N$, the summations over $n$ in \eref{eq:density2} can be approximated by integrals over $u$. This leads to 
\begin{align}
\label{eq:rho+s}
    \bar{\rho}^+(z) &\simeq \frac{1}{2} N^{\frac{1}{\alpha+2}} \int_0^1 \phi_{\frac{Nu}{2}}^2(N^{\frac{1}{\alpha+2}}z) \, du,\\[5mm]
    \label{eq:rho-s}
\bar{\rho}^-(z) &\simeq \frac{1}{2} N^{\frac{1}{\alpha+2}}\int_0^1  \chi_{\frac{Nu}{2}}^2(N^{\frac{1}{\alpha+2}}z)\, du. 
\end{align}
The eigenfunctions $\phi_M(N^{\frac{1}{\alpha+2}}z)$ and $\chi_M(N^{\frac{1}{\alpha+2}}z)$ are given in Eqs.~\eqref{eq:phin} and \eqref{eq:chin} respectively, in terms of generalized Laguerre polynomials $L_M^{(\pm\gamma)}(4 M w)$ with $\gamma=(\alpha+1)/\alpha+2)$, $M=Nu/2$, and $w=\mu |z|^{\alpha+2}/[(\alpha+2)u]$. For large $M$, we can use the Plancherel-Rotach asymptotics for Laguerre polynomials~\cite{szeg1939orthogonal}, 
\begin{equation}
    L_M^{(\beta)} (4 M w) \simeq \frac{(-1)^M}{\sqrt{2\pi M}\, 2^{\beta}} \, \frac{e^{2 M w}\, g_M(w)}{w^{\beta/2} [w (1-w)]^{1/4}}\,, 
    \label{eq:Plancherel-Rotach}
\end{equation}
$0<w<1$, where 
\begin{equation}
\begin{split}
    g_N(w)& =\sin\left[h_N(w)\right]+ O(1/N), \quad \text{with} \\ h_N(w) & = 2 N\left(\sqrt{w(1-w)}-\cos^{-1}\sqrt{w}\right)+\frac{3\pi}{4}\, .  
    \label{eq:gN}
\end{split}
\end{equation}
Using Stirling's approximation,
\begin{equation}
   \frac{\Gamma(M +1)}{\Gamma(M + 1 \mp\gamma) }\simeq M^{\pm \gamma} \quad\text{for large } M, 
\end{equation}
we find that both integrals in Eqs.~\eqref{eq:rho+s} and \eqref{eq:rho-s} yield the same contribution,
\begin{equation}
  \bar{\rho}^\pm(z) \simeq   \frac{\sqrt{\mu(\alpha+2)}}{2\pi}\, |z|^{\alpha/2}\, \int_{\frac{\mu|z|^{\alpha+2}}{\alpha+2}}^1\, \frac{g_{\frac{Nu}{2}}^2\left(\frac{\mu|z|^{\alpha+2}}{(\alpha+2)u}\right)}{\sqrt{u-\frac{\mu|z|^{\alpha+2}}{\alpha+2}}}\, du\, .
  \label{eq:rho-int}
\end{equation}
To the leading order, $g_{Nu/2}^2(x)\simeq 1/2$, while the remaining cosine term oscillates highly with respect to $y$ for large $N$. Therefore, using the leading term and performing the integral, we get 
\begin{equation}
\bar{\rho}^\pm (z) \simeq \frac{\sqrt{\mu(\alpha+2)}}{2\pi}\, |z|^{\alpha/2}\, \sqrt{1-\frac{\mu|z|^{\alpha+2}}{\alpha+2}} . 
\label{eq:rho-pm-scal}
\end{equation}
Therefore, to summarize, the full density profile $\rho_N(x)$ admits the scaling form given in \eref{eq:rho-scaling}, where the scaling function $\bar{\rho}(z)=\bar{\rho}^+(z) + \bar{\rho}^-(z)$ is obtained by using \eref{eq:rho-pm-scal}, as 
\begin{equation}
\bar{\rho} (z) \simeq \frac{\sqrt{\mu(\alpha+2)}}{\pi}\, |z|^{\alpha/2}\, \sqrt{1-\frac{\mu|z|^{\alpha+2}}{\alpha+2}} . 
\label{eq:rho-scal}
\end{equation}
For $\alpha=0$, \eref{eq:rho-scal} reduces to
\begin{equation}
    \bar{\rho}(z) = \frac{\sqrt{2\mu}}{\pi}\, \sqrt{1-\frac{\mu z^2}{2}},
    \label{eq:Wigner}
\end{equation}
 which is in agreement with \eref{wigner}. In \fref{fig:fig2}, we plot $N^{\frac{1}{\alpha+2}} \rho_N(N^{\frac{1}{\alpha+2}}z)$ as a function of $z$ [see~\eref{eq:rho-scaling}], obtained by using the expression of $\rho_N(x)=\rho_N^+(x)+\rho_N^-(x)$ from Eqs.~\eqref{eq:KN2a1} and \eqref{eq:KN2a2} in \eref{eq:rho-scaling}, for $N=100$, $\mu=1$ and different values of $\alpha$, and compare it with the scaling function $\bar{\rho}(z)$ given in \eref{eq:rho-scal}, and find excellent agreement away from the origin. As discussed earlier [see after \eref{eq:density0}], $N^{\frac{1}{\alpha+2}} \rho_N(N^{\frac{1}{\alpha+2}}z)$, for large $N$,  converges to $\bar\rho_N(z)$ in \eref{eq:rho-scal} as $N^{-\frac{\alpha+1}{\alpha+2}}$.

We would like to point out an interesting contrast between the density profiles for $\alpha=0$ and $\alpha>0$. 
For $\alpha=0$, the density profile
is given by the Wigner semi-circular law in \eqref{eq:Wigner}, which is maximal at $z=0$. However, for
any $\alpha>0$, the average density profile has a minimum at $z=0$, indicating a `depletion'
zone at $z=0$ (see Fig.~\ref{fig:fig2}).  This depletion zone near the origin emerges from the suppression of the quantum probabilities $|\psi_n(x)|^2$ for the excited states ($n>0$) near the origin for $\alpha>0$ (see \fref{fig:eigenfunctions}) and the density is a arithmetic mean of the these quantum probabilities, $\rho_N(x)= N^{-1} \sum_{n=0}^{N-1} |\psi_n(x)|^2$, as given in \eref{eq:kxx}.

\subsection{Scaling behavior of the kernel}
\label{s:kernel}

In this section, we study the scaling behavior of the correlation kernel $K_N(x,y)$ given in Eqs.~\eqref{eq:kernel-MB2}--\eqref{eq:K-},  when (i) both $x$ and  $y$ are in the bulk of the density profile, (ii) near its rightmost edge, and (iii) close to the origin. 

\subsubsection{Bulk kernel}

We first consider the case when both $x$ and $y$ are in the bulk of the density profile (see \fref{fig:fig2}). Since the density profile is symmetric around the origin, for simplicity, we consider the positive side, i.e., $x>0$ and $y>0$.
In the bulk, the typical interparticle separation $\ell(x)\sim [N\rho_N(x)]^{-1}$. On the other hand, $\rho_N(x)$ itself scales with $N$ as given in \eref{eq:rho-scaling}. Therefore, $\ell(x)$ scales with $N$ as $N^{-\gamma}$ with $\gamma=(\alpha+1)/(\alpha+2)$. This suggests the scaling form of the bulk kernel
\begin{equation}
    K_N(x,y) = \frac{1}{N^{-\gamma}}\,  F\left(\frac{|x-y|}{N^{-\gamma}},\frac{x}{N^{\frac{1}{\alpha+2}}}\right).
    \label{eq:kernel-bulk-scaling}
\end{equation}

To obtain the scaling function $F(w,z)$ in \eref{eq:kernel-bulk-scaling}, we analyze the kernel $K_N(x,y)$ in the scaling limit 
\begin{equation}
   F(w,z) =\lim_{N\to\infty} N^{-\gamma} \, \frac{1}{N}\, K_N\bigl(N^\frac{1}{\alpha+2} z, N^\frac{1}{\alpha+2} z+ N^{-\gamma} w\bigr).
   \label{eq:kernel-bulk-scal2}
\end{equation}
for $z>0$. 

Similar to the calculation of the scaling function of the density profile in \sref{s:density}, here, we follow the three main steps:
First, we use Eqs.~\eqref{eq:kernel-MB2}--\eqref{eq:K-} in \eref{eq:kernel-bulk-scal2}. Next, we convert the summation over $n$ to an integral over $u$ by making a change of variable $n=Nu/2$ for large $N$. Finally, we employ the Plancherel-Rotach asymptotic given by
\eref{eq:Plancherel-Rotach}, to arrive at
\begin{align}
  F(w,z)&=\lim_{N\to\infty}  \frac{\sqrt{\mu(\alpha+2)}}{\pi}\, z^{\alpha/2}\, \int_{\frac{\mu \,z^{\alpha+2}}{\alpha+2}}^1\, \frac{du}{\sqrt{u-\frac{\mu z^{\alpha+2}}{\alpha+2}}}\notag\\
  &\times
  \, g_{\frac{Nu}{2}}\left(\frac{\mu \,z^{\alpha+2}}{(\alpha+2)u}\right)\, g_{\frac{Nu}{2}}\left(\frac{\mu \,(z+w/N)^{\alpha+2}}{(\alpha+2)u} \right).
  \label{eq:bulk-int}
\end{align}

From \eref{eq:gN}, we get
\begin{align}
    g_N(z)\, g_N(z+\epsilon) &= \frac{1}{2}\bigl(\cos[h_N(z)-h_N(z+\epsilon)] \notag\\
    &- \cos[h_N(z)+h_N(z+\epsilon)]\bigr) \notag\\
    &\simeq \frac{1}{2} \left(\cos[-\epsilon h'_N(z)] -\cos[2h_N(z)]\,  \right).
    \label{eq:gg}
\end{align}
The second cosine term in \eref{eq:gg} oscillates highly for large $N$, and therefore, contributes only sub-dominantly.
Using only the first term of \eref{eq:gg} in \eref{eq:bulk-int}, we get
\begin{align}
F(w,z)& =\frac{\sqrt{\mu(\alpha+2)}}{2\pi}\, z^{\alpha/2}\, \int_{\frac{\mu \,z^{\alpha+2}}{\alpha+2}}^1\, \frac{du}{\sqrt{u-\frac{\mu  z^{\alpha +2}}{\alpha +2}}}\notag\\
&\times \cos \left(w \sqrt{\mu\,(\alpha +2)  } \,z^{\alpha /2} \sqrt{u-\frac{\mu  z^{\alpha +2}}{\alpha +2}}\right).
\label{eq:fwz-1}
\end{align}
The integral in \eref{eq:fwz-1} can be performed exactly, yielding
\begin{equation}
    F(w,z) = \frac{\sin\left(\pi w \bar{\rho}(z)\right)}{\pi w}\, ,
    \label{eq:fwz}
\end{equation}
where $\bar\rho(z)$ is the scaling function of the density profile given in \eref{eq:rho-scal}.
It follows from Eqs.~\eqref{eq:kernel-bulk-scaling} and \eqref{eq:fwz} that the correlation kernel has the scaling form
\begin{equation}
    K_N(x+x', x+y') = \frac{1}{\ell(x)}\, K_\mathrm{bulk} \left(\frac{|x'-y'|}{\ell(x)}\right), 
\end{equation}
where 
\begin{equation}
    \ell(x)= \frac{1}{N \rho_N(x)}\,
\end{equation}
and $K_\mathrm{bulk}(w)$ is given by the sine kernel~\cite{mehta2004random,forrester2010log,Eisler2013,Dean2019review},
\begin{equation}
    K_\mathrm{bulk}(w) = \frac{\sin \pi w}{\pi w}\, .
\end{equation}

It is easy to check that $K_\text{bulk}(w)\to 1$ in the limit $w\to 0$, yielding $K_N(x,x)=N\rho_N(x)$, as expected.

\subsubsection{Edge kernel}

Here, we consider the case when both $x$ and $y$ are near one of the edges, say the right edge, denoted by $x_{\text{edge}}$. Before proceeding further, we discuss the expressions for $x_{\text{edge}}$ and the interparticle distance denoted by $W_N$. From Eqs~\eqref{eq:rho-scaling} and \eqref{eq:rho-scal}, it is easy to see that the support of the density vanishes at  
\begin{equation}
x_{\text{edge}} = z_{\text{edge}} \,N^{\frac{1}{\alpha+2}}\quad \text{with}\quad z_{\text{edge}} = \left(\frac{\alpha+2}{\mu}\right)^{\frac{1}{\alpha+2}}\,.
\end{equation}
The interparticle distance $W_N$ at the edge can be estimated by
\begin{equation}
\label{eq:xedge}
\int_{x_\text{edge}-W_N}^{x_{\text{edge}}} \rho_N(x)\,dx \approx \frac{1}{N}\,. 
\end{equation}
By using the scaled variables $z=x/N^{\frac{1}{\alpha+2}}$, we get
\begin{equation}
\label{eq:zedge}
    \int_{z_\text{edge}-\frac{W_N}{N^{\frac{1}{\alpha+2}}}}^{z_{\text{edge}}} \bar{\rho}(z)\,dz \approx \frac{1}{N}\,. 
\end{equation}
To estimate $W_N$ (especially, its scaling with $N$), we probe short length scales around $z_\text{edge}$. Taking $z=z_\text{edge}-u$, Eq.~\eqref{eq:rho-scal} gives 
\begin{equation}
\bar{\rho}(z) \approx \frac{\sqrt{\mu(\alpha+2)}}{\pi} z_\text{edge}^{\alpha/2}\sqrt{\frac{\alpha+2}{z_\text{edge}}} \sqrt{u}\,.
\label{eq:rho_exp}
\end{equation}
Equation~\eqref{eq:rho_exp} along with 
\eqref{eq:zedge} gives 
\begin{equation}
\label{eq:rho_exp1}
\frac{\sqrt{\mu(\alpha+2)}}{\pi}\sqrt{\alpha+2}\,\,z_\text{edge}^{\frac{\alpha-1}{2} }\,
\int_0^{W_N/N^{\frac{1}{\alpha+2}}} du \sqrt{u}  \approx \frac{1}{N}\,.
\end{equation}
\Eref{eq:rho_exp1} immediately gives the estimate for $W_N$ as
\begin{equation}
    W_N = \frac{z_\text{edge}}{\alpha+2}\, N^{-\frac{2\alpha+1}{3(\alpha+2)}}, 
    \label{eq:WN}
\end{equation}
where we have omitted a multiplicative factor of $(3\pi/2)^{2/3}$ for convenience later. Therefore, near the right edge, we expect the scaling form 
\begin{equation}
    K_N(x,y)= \frac{1}{W_N} \, K_\text{edge} \left(\frac{x-x_\text{edge}}{W_N}, \frac{y-x_\text{edge}}{W_N}\right).
    \label{eq:KNAiry}
\end{equation}
The scaling function can be found by evaluating
\begin{equation}
    K_\text{edge}(u,v)=\lim_{N\to\infty} \, W_N\, K_N(x_\text{edge}+W_N\, u, x_\text{edge}+W_N \,v)\,,
    \label{eq:Kuv}
\end{equation}
where $K_N(x,y)$ is given in \eref{eq:kernel-MB2}, together with Eqs.~\eqref{eq:K+s}--\eqref{eq:KN2}. We note that the variables $x$ and $y$ in the expression of $K_N(x,y)$ appear through the combination
\begin{equation}
    z_1 = \frac{2\mu\, x^{\alpha+2}}{\alpha+2} \quad\text{and}\quad
    z_2 = \frac{2\mu\, y^{\alpha+2}}{\alpha+2}\, .
    \label{eq:z1z20}
\end{equation}
Substituting $x=x_\text{edge}+W_N\, u$ and $y=x_\text{edge}+W_N\, v$, we have 
\begin{equation}
    z_1 = 2N + 2 N^{1/3}\,u\, ,\quad\text{and}\quad
     z_2 = 2N + 2 N^{1/3}\,v\, .
      \label{eq:z1z2}
\end{equation}
Using \eref{eq:z1z2} in Eqs.~\eqref{eq:KN1}--\eqref{eq:KN2}, we get (see \aref{a:edge})
\begin{equation}
    K_\text{edge} (u,v)= \frac{\text{Ai}(u)\text{Ai}'(v) - \text{Ai}'(u)\text{Ai}(v)}{u-v}\, 
    \label{eq:AiryK}
\end{equation}
where $\text{Ai}(u)$ and $\text{Ai}'(u)$ are the Airy and AiryPrime functions respectively.

\Eref{eq:KNAiry} along with \eref{eq:AiryK} gives the kernel near the right edge. Taking the limit $u\to v$ in \eref{eq:AiryK} gives the density profile near the right edge. From \eref{eq:AiryK}, we have
\begin{equation}
    \lim_{u\to v} K_\text{edge} (u,v)\equiv R_\text{edge}(v)=\left[\text{Ai}'(v)\right]^2 - v\left[\text{Ai}(v)\right]^2, 
    \label{eq:R}
\end{equation}
where we have used the Airy equation $\text{Ai}''(v) = v\, \text{Ai}(v)$. Therefore, the density profile near the right edge is given by 
\begin{equation}
    \rho_N^\text{edge}(x) = \frac{1}{NW_N}\, R_\text{edge}\left(\frac{x-x_\text{edge}}{W_N}\right),
    \label{eq:edge-density}
\end{equation}
where $R_\text{edge}(v)$ is given in \eref{eq:R}. The edge density $\rho_N^\text{edge}(x)$, in the $v\to-\infty$ limit of $R_\text{edge}(v)$ in \eref{eq:edge-density}, is expected to  coincide with bulk density $\rho_N(x)$ in \eref{eq:rho-scaling}, with $\bar\rho(z)$ given by the edge limit in \eref{eq:rho_exp}. Indeed, using the asymptotic limit $R(v) \sim \sqrt{-v}/\pi$ in \eref{eq:edge-density}, we find this precise agreement.

\subsubsection{Kernel close to the origin}
\label{s:korigin}

To analyze the kernel $K_N(x,y)$ near the origin, we first estimate the typical interparticle separation $\ell_0$ near the origin, using  $\int_0^{\ell_0} \rho_N(x)\, dx \approx 1/N$, or equivalently, 
\begin{equation}
    \int_0^{\ell_0/N^{1/(\alpha+2)}} \bar{\rho}(z)\, dz \approx \frac{1}{N}\, .
    \label{eq:sepO}
\end{equation}
Close to the origin, say on the positive side, from \eref{eq:rho-scal}, using 
\begin{equation}
    \bar{\rho}(z) \simeq \frac{\sqrt{\mu (\alpha+2)}}{\pi}\, z^{\alpha+2},
    \label{eq:rhoz-0}
\end{equation}
 in \eref{eq:sepO}, we get
\begin{equation}
    \ell_0 = \left(\frac{\alpha+2}{\mu}\right)^{\frac{1}{\alpha+2}}\, N^{-\frac{1}{\alpha+2}},
\end{equation}
where we have ignored a multiplicative factor for later convenience. Therefore, near the origin, on the positive size, we expect the scaling form of the correlation kernel as
\begin{equation}
    K_N(x,y)=\frac{1}{\ell_0}\, K_0\left(\frac{x}{\ell_0}, \frac{y}{\ell_0}\right)\, .
    \label{eq:k-o}
\end{equation}
The scaling function can be found by evaluating
\begin{equation}
    K_0(u,v)=\lim_{N\to\infty} \, \ell_0\, K_N(\ell_0\, u, \ell_0 \,v)\,,
    \label{eq:Kuv2}
\end{equation}
where $K_N(x,y)$ is given in \eref{eq:kernel-MB2}, together with Eqs.~\eqref{eq:K+s}--\eqref{eq:KN2}. Since the variables $x$ and $y$ in the expression of $K_N(x,y)$ appear through the combination in \eref{eq:z1z20}, substituting $x=\ell_0 \,u$ and $y=\ell_0\, v$, we get $z_1=(2/N)\, u^{\alpha+2}$ and $z_2=(2/N)\, v^{\alpha+2}$ in Eqs.~\eqref{eq:KN1} and \eqref{eq:KN2} with $M=N/2$ (for simplicity, we take $N$ even; this is immaterial in the large-$N$ limit). As shown in \aref{a:kno}, we get
\begin{equation}
    K_0(u,v) = K_0^+ (u,v)+ K_0^-(u,v),
    \label{eq:k0pm}
\end{equation}
where $K_0^\pm(u,v)$ are given by
\begin{align}
    K_0^\pm(u,v) & = \frac{\alpha+2}{2}\, \frac{(\tilde{z}_1\, \tilde{z}_2)^{\gamma/2}}{\tilde{z}_1-\tilde{z}_2}\, \notag\\
    &\times \Bigl[\sqrt{\tilde{z}_1} J_{\mp\gamma }\left(2 \sqrt{\tilde{z}_2}\right) J_{\mp\gamma +1}\left(2 \sqrt{\tilde{z}_1}\right) 
    \notag \\ &-\sqrt{\tilde{z}_2} J_{\mp\gamma }\left(2 \sqrt{\tilde{z}_1}\right) J_{\mp\gamma +1}\left(2 \sqrt{\tilde{z}_2}\right)\Bigr],
    \label{eq:k0pm1}
\end{align}
where 
\begin{equation}
   \tilde{z}_1=u^{\alpha+2}, \quad \tilde{z}_2=v^{\alpha+2}, \quad\text{and}\quad
   \gamma =  \frac{\alpha+1}{\alpha+2}\,.
   \label{eq:tilde_z}
\end{equation}
\Eref{eq:k-o} along with Eqs.~\eqref{eq:k0pm} and \eqref{eq:k0pm1} gives the kernel near the origin. 

The density close to the origin can be obtained from the kernel $K_0^\pm(u,v)$ in \eref{eq:k0pm1},  by taking the limit $u\to v$, which gives
\begin{align}
   &  \lim_{u\to v} K_0^\pm(u,v) \equiv R_0^\pm(v) = \frac{\alpha+2}{2}\, \tilde{z}_2^\gamma\,\Biggl[
    J^2_{\mp\gamma +1}\left(2 \sqrt{\tilde z_2}\right)
      \notag\\
     &
    +J_{\mp\gamma }\left(2 \sqrt{\tilde z_2}\right) \left(\frac{J_{\mp\gamma +1}\left(2 \sqrt{\tilde z_2}\right)}{\sqrt{\tilde z_2}}-J_{\mp\gamma +2}\left(2 \sqrt{\tilde z_2}\right)\right)
    \Bigg].
    \label{eq:k0vv}
\end{align}\
The density near the origin, using Eqs.~\eqref{eq:k-o} and \eqref{eq:k0vv}, is given by
\begin{equation}
    \rho_N(y) = \lim_{x\to y} \frac{1}{N} K_N(x,y)  =\frac{1}{N\ell_0} \left[R_0^+\left(\frac{y}{\ell_0}\right) + R_0^-\left(\frac{y}{\ell_0}\right)\right].
    \label{eq:rho-0}
\end{equation}
For large $v$, we have
\begin{equation}
    R_0^\pm (v)  \simeq \frac{\alpha+2}{2 \pi}\, v^{\alpha/2}.
\end{equation}
Therefore, \eref{eq:rho-0} yields
\begin{equation}
    \rho_N(x) \simeq \frac{1}{\sqrt{N}}\, \frac{\sqrt{\mu (\alpha+2)}}{\pi }\, x^{\alpha/2}\quad\text{as ~} x\to 0.
\end{equation}
This is in perfect agreement with the $x\to 0$ limit of the bulk density $\rho_N(x)$ in \eref{eq:rho-scaling}, where the limiting scaling function  $\bar\rho(z)$ is given by \eref{eq:rhoz-0}.

We finally remark that the expression $K^{\pm}_0(u,v)$ in \eref{eq:k0pm1} resembles the well-known Bessel kernel in random matrix theory after a suitable transformation. To see the connection to the Bessel kernel more clearly, we use the following identity satisfied by the Bessel functions~\cite{abramowitz1965handbook},
\begin{equation}
    J_{\nu + 1}(\sqrt{x}) = \frac{\nu}{\sqrt{x}}J_\nu(\sqrt{x}) - J'_\nu(\sqrt{x})\,,
    \label{eq:bessel-id}
\end{equation}
where $J'_\nu(z) = \frac{\dd J_\nu(z)}{\dd z}$. Using \eref{eq:bessel-id} in Eqs.~\eqref{eq:k0pm1}--\eqref{eq:tilde_z}, after simplifications, we get
\begin{equation}\label{eq:Kmb}
    K_0^\pm(u,v) = 2(\alpha+2) (u\,v)^{\frac{\alpha+1}{2}} K_{\rm Bessel}^{(\mp\gamma)}(4 u^{\alpha+2}, 4 v^{\alpha+2})\,.
\end{equation}
where the Bessel kernel is given by \cite{forrester2010log, PJF93}
\begin{equation}
    K_{\rm Bessel}^{(\nu)}(x,y) =\frac{\sqrt{y}\,J'_\nu(\sqrt{y})J_\nu(\sqrt{x}) - \sqrt{x}\,J'_\nu(\sqrt{x})J_\nu(\sqrt{y})}{2(x-y)}\,.
\end{equation}
To make the identification with the Bessel kernel more precise. We recall how a kernel transforms under a change of variables. Suppose we have a kernel $K(u,v) = \sum_{n=0}^{N-1} \psi_n(u) \psi_n(v)$ where $\psi_n(u)$'s are normalized single particle eigenfunctions. Suppose now we make a transformation
\begin{equation}\label{eq:chagev2}
    u \to u'=f(u)\quad \text{and} \quad v \to v'=f(v)
\end{equation}
where $f(u)$  is an arbitrary function. Then
\begin{equation}\label{eq:changek}
    K(u,v) \to \sqrt{|f'(u) f'(v)|}\tilde K(f(u),f(v))
\end{equation}
where $\tilde K(u',v')$ is the transformed kernel. Eq.~\eqref{eq:changek} follows from the transform of wavefunction $\psi_n(u) \to \sqrt{f'(u)}\tilde \psi(u')$ which preserves its normalization. Applying this transformation, with the choice $f(u) = 4u^{\alpha+2}$, we see that 
\begin{equation}\label{eq:Km_change}
    K^\pm_0(u,v) \to 4(\alpha+2) (u\, v)^{\frac{\alpha+1}{2}}\tilde K_0^\pm(4u^{\alpha+2}, 4u^{\alpha+2})\,. 
\end{equation}
Comparing the right hand sides of Eq.~\eqref{eq:Kmb} and Eq.~\eqref{eq:Km_change} we see that the transformed kernel reads
\begin{equation}\label{eq:KhalfB}
    \tilde K_0^\pm(u',v') = \frac{1}{2}K_{\rm Bessel}^{(\mp\gamma)}(u',v')\,.
\end{equation}
Therefore, the total kernel in Eq.~\eqref{eq:k0pm} in the transformed coordinates becomes
\begin{equation}
    \tilde K_0(u',v') = \frac{1}{2}\left[K_{\rm Bessel}^{(\gamma)}(u',v') + K_{\rm Bessel}^{(-\gamma)}(u',v')\right]\,,
    \label{eq:kedge_bessel}
\end{equation}
where recall that $\gamma= (\alpha+1)/(\alpha+2)$. \Eref{eq:kedge_bessel} is a sum of two Bessel kernels with different indices $\pm \gamma$, as opposed to a single Bessel kernel. Thus, the sluggish fermions in the ground state form a DPP with a `new kernel' that has not been encountered before in the fermionic literature, to the best of our knowledge. This is one of the important findings of our paper.

\subsection{JPDF in the half-space}
\label{s:JPDF-half}

It is useful and important to examine the structure of the many-body ground state in a setting where the JPDF can be obtained in an explicit form. For noninteracting fermions, the JPDF is determined by the Slater determinant constructed from the single-particle eigenfunctions, and in general, it can be expressed in terms of the determinant of the kernel. However, obtaining a closed analytical form of the JPDF in the full space is difficult because of the complicated structure of the eigenfunctions. A considerable simplification occurs when the system is restricted to a half-space with an appropriate boundary condition at the origin. In this case, the parity structure of the eigenfunctions allows one to construct the many-body ground state solely from either the odd or the even single-particle eigenstates, corresponding to Dirichlet or Neumann boundary conditions, respectively. As we show below, this restriction leads to a remarkably simple and elegant form of the ground-state wavefunction and consequently of the JPDF.

The JPDF $P(x_1, x_2,\dotsc, x_N)$ of the positions of the noninteracting fermions in the presence of an external trap $V_\mathrm{ext}(x)=\frac{1}{2}m_\mathrm{eff}\,\omega^2 |x|^{\alpha+2}$ is given by \eref{eq:JPDF-MB} in terms of the determinants of the kernel $K_N(x,y)$ given in Eqs.~\eqref{eq:kernel-MB2}--\eqref{eq:K-} and \eqref{eq:K+s}--\eqref{eq:KN2}. As mentioned above, it is generally difficult to obtain an explicit form and interestingly, it turns out that one can obtain an explicit form of the JPDF on the positive (or equivalently negative) half space with either Dirichlet or Neumann boundary conditions at the origin.

The many-body ground state wavefunction vanishes at the origin for the 
Dirichlet boundary condition. Therefore, the $\psi_0(x_1, x_2, \dotsc, x_N)$ in \eref{eq:MB-psi} must be constructed out of the odd eigenfunctions $\psi_{2n+1}(x)=\chi_n(x)$, with $n=0, 1, \dotsc, N-1$, where $\chi_n(x)$ is given in \eref{eq:chin}. Since the terms with lower powers in the associated Laguerre polynomials in the Slater determinant can be removed by row or column operations, we can retain only the highest power in $\chi_n(x)$. Moreover,  the lead coefficients can be absorbed in the normalization. Therefore, from \eref{eq:chin}, we have
\begin{equation}
    \chi_n(x) \propto x^{\alpha+1}\, e^{-\frac{\mu |x|^{\alpha+2}} {\alpha+2}}\, x^{n(\alpha+2)}.
    \label{eq:chin-lead}
\end{equation}
Using \eref{eq:chin-lead} in \eref{eq:MB-psi}, we obtain
\begin{align}
    \psi_0(x_1, x_2, \dotsc, x_N) &=A  e^{-\frac{\mu } {\alpha+2} \sum_{i=1}^N x_i^{\alpha+2}} \notag \\&\times \prod_{i=1}^N x_i^{\alpha+1} \,\prod_{1\le i <j \le N} (x_i^{\alpha+2}-x_j^{\alpha+2}),
\end{align}
where $A$ is a normalization constant. Therefore, the JPDF in \eref{eq:JPDF1} is given by
\begin{align}
\label{eq:jpdfm}
    P^-(x_1, x_2, &\dotsc, x_N) = \mathcal{N}_- \, e^{-\frac{2 \mu } {\alpha+2} \sum_{i=1}^N x_i^{\alpha+2}} \notag \\&\times \prod_{i=1}^N x_i^{2(\alpha+1)} \,\prod_{1\le i <j \le N} (x_i^{\alpha+2}-x_j^{\alpha+2})^2,
\end{align}
where $\mathcal{N}_-$ is the normalization constant.

Similarly, the many-body ground state wavefunction for the Neumann boundary condition can be constructed using the even eigenfunctions $\psi_{2n}=\phi_n(x)$ in the Slater determinant, yielding
\begin{align}
\label{eq:jpdfp}
    P^+(x_1, x_2,  \dotsc, x_N) = & \mathcal{N}_+ e^{-\frac{2 \mu } {\alpha+2} \sum_{i=1}^N x_i^{\alpha+2}} \notag \\&\times \prod_{1\le i <j \le N} (x_i^{\alpha+2}-x_j^{\alpha+2})^2,
\end{align}
where $\mathcal{N}_+$ is the normalization constant.

The explicit form of the JPDF [Eqs.~\eqref{eq:jpdfm} and ~\eqref{eq:jpdfp}] obtained in the half-space is useful for several reasons. First, it reveals the underlying Vandermonde-type structure of the fermionic correlations, thereby establishing a direct connection with log-gas type models and random-matrix ensembles after a suitable change of variables. Such representations provide valuable intuition about the effective repulsion between fermions and the role of the confining potential and  spatially varying effective mass in shaping the equilibrium configuration. Second, the explicit JPDF serves as a convenient starting point for computing various statistical properties of the fermion positions, such as density profiles, gap probabilities, and extreme-value statistics in finite and the large-$N$ limit. 

We elaborate the above points in further detail here. A useful insight follows from the change of variables 
$y_i=x_i^{\alpha+2}$, which maps the JPDF obtained [Eqs.~\eqref{eq:jpdfm} and ~\eqref{eq:jpdfp}] to a standard 
log-gas form. Using $dx_i=\frac{1}{\alpha+2}y_i^{-\gamma}dy_i$ [where recall $\gamma = (\alpha+1)/(\alpha+2)$],  
one finds that the Dirichlet JPDF [Eq.~\eqref{eq:jpdfm}] transforms as
\begin{align}
P^-(y_1,\dots ,y_N)\propto
\prod_{1\le i<j\le N}(y_i-y_j)^2
\prod_{i=1}^{N}y_i^{\gamma}\,
e^{-\frac{2\mu}{\alpha+2}y_i}.
\end{align}
This is precisely the joint eigenvalue distribution of the Laguerre (Wishart) random-matrix ensemble~\cite{PJF93, forrester2010log} with Dyson index $\beta=2$ and parameter 
$\gamma=(\alpha+1)/(\alpha+2)$.

Similarly, for the Neumann boundary condition, from the JPDF given in Eq.~\eqref{eq:jpdfp} one obtains
\begin{align}
P^+(y_1,\dots ,y_N)\propto
\prod_{1\le i<j\le N}(y_i-y_j)^2
\prod_{i=1}^{N}y_i^{-\gamma}\,
e^{-\frac{2\mu}{\alpha+2}y_i},
\end{align}
corresponding to a Laguerre ensemble with parameter 
$-\gamma$. This mapping establishes a direct correspondence 
between the statistics of fermion positions in the sluggish quantum system 
and the eigenvalue statistics of Laguerre random matrices. Consequently, as indicated earlier, 
many properties of the fermion gas—such as density profiles, gap probabilities, 
and extreme-value statistics can be analyzed using the well-developed 
framework of Laguerre ensembles and associated determinantal point processes.

\section{Conclusion and outlook}
\label{s:conclusion}

In this work, we studied a class of one-dimensional quantum systems characterized by a spatially varying kinetic term arising from the continuum limit of an inhomogeneous tight-binding model with bond-dependent hopping amplitudes. In this limit, the lattice model maps onto a Schr\"odinger equation with an effective mass scaling as $m_\mathrm{eff}(x)= m_\mathrm{eff}\, |x|^{\alpha}$. This construction provides a concrete microscopic route from engineered optical lattices with a position-dependent tunneling to a continuum theory with a smoothly varying effective mass. The resulting dynamics, which we termed as \emph{sluggish quantum mechanics}, is governed by a progressive suppression of kinetic motion at large distances.

For the case with no external confinement, we obtained exact analytical expressions for the eigenfunctions and the full quantum propagator. The spatially varying effective mass qualitatively modifies the structure of plane-wave eigenstates, leading to nontrivial spatial modulation and, subsequently, to anomalous spreading of an initial wavefunction over time. Therefore, even in the absence of an external potential, the power-law variation of the effective mass fundamentally alters long-distance behavior compared to conventional free-particle dynamics with a space-independent mass. In the presence of the confining potential $V_\mathrm{ext}(x) = \frac{1}{2}m_\mathrm{eff}(x)\omega^2 x^2 =\frac{1}{2} m_\mathrm{eff}\, \omega^2 |x|^{\alpha+2}$, we demonstrated that the model remains exactly solvable. We derived the complete spectral properties and closed-form expressions for the propagator. The connection between the power-law effective mass function and the form of the confining potential ensures analytical tractability and generalizes the harmonic oscillator structure ($\alpha=0$) to $\alpha >0$.

We then investigated the many-body ground state of $N$ noninteracting spinless fermions in the confining potential $V_\mathrm{ext}(x) = \frac{1}{2} m_\mathrm{eff}\, \omega^2 |x|^{\alpha+2}$. Using the exact single-particle eigenfunctions, we constructed the ground-state wavefunction and obtained the joint probability density function. The associated quantum correlation kernel was derived exactly. This enabled us to compute the density profile both at finite $N$ and in the large-$N$ limit. We analyzed the universal scaling behavior of the kernel in the bulk, at the edge, and near the origin. Our results generalize the well-known results on harmonically trapped fermions to a system with a spatially varying effective mass.

We would like to remark that
qualitatively new physics appears near $x=0$, where a new kernel in~\eref{eq:kedge_bessel} describes
the underlying `determinantal point process' (DPP) near the origin. We find that the kernel near the origin
is a sum of two Bessel kernels with different 
indices, as opposed to a single Bessel kernel that appears in Laguerre ensembles~\cite{ PJF93, forrester2010log, nadal2009}. Thus, the sluggish fermions in the ground state form a DPP characterized by a novel kernel that, to the best of our knowledge, has not appeared previously in the fermionic literature. This represents one of the key findings of our work.

Several open directions and natural extensions emerge from our work. 
While we have focused on a specific power-law profile, the continuum construction from inhomogeneous tight-binding models suggests a broader class of spatially varying kinetic terms. It would be interesting to classify which functional forms of $D(x)$ admit exact solvability and whether additional families lead to integrable many-body structures or determinantal point processes with modified kernels~\cite{GUHR1998189,mehta2004random,forrester2010log,oxford_book}. Needless to mention, the dynamical properties of sluggish quantum systems beyond the quantum propagator merit further investigation. Questions related to quantum quenches~\cite{mitra_cmp_review}, wave-packet spreading, entanglement growth~\cite{SKD_2024,SK24}, and return probabilities~\cite{GORIN200633} could reveal novel dynamical scaling regimes induced purely by kinetic inhomogeneity~\cite{Gora1969,Bastard1975,Zhu1983,Roos1983,harrison2016quantum,osti_5095386}. Furthermore, recent works probe the appearance of Wishart-Laguerre RMT statistics in two-dimensional fermions~\cite{dejong2025,dixmerias2025} directly in the continuum. This naturally raises the question whether a $d$-dimensional sluggish quantum mechanics is (i) solvable and (ii) experimentally accessible, even at the single particle level~\cite{maltezopoulos2003, verstraten2025}.
Another interesting direction is to generalize our work to Fermions at finite temperature~\cite{Dean2016,grela17,KfiniteT} where smearing effects in density profile and other quantities can be thoroughly investigated. 
One can also study the bosonic version of the problem and investigate Bose-Einstein condensation and other aspects~\cite{kms2025_bosons} in the sluggish case. 

Our work is directly motivated by engineered ultracold-atom systems in optical lattices, where spatially varying tunneling amplitudes can be realized using programmable potentials. This raises the possibility of experimentally probing sluggish quantum dynamics, density profiles, and edge scaling in controlled setups. In fact, our work put forwards a route to potentially realize the Laguerre-Wishart ensemble~\cite{PJF93, forrester2010log} in experiments~\cite{dixmerias2025}. More generally, the continuum limit of inhomogeneous lattice models provides a versatile and physically motivated route to quantum systems with spatially varying effective mass. The exact solvability uncovered here demonstrates that such systems are not only experimentally relevant but also analytically tractable, opening new avenues for exploring the interplay between kinetic inhomogeneity, external confinement, and many-body quantum correlations.

\section*{Acknowledgment}
We acknowledge the support from the Science and Engineering Research Board (SERB, Government of India), under the VAJRA faculty scheme (VJR/2017/000110 and VJR/2019/00007). MK acknowledges support from the Department of Atomic Energy, Government of India, under project no. RTI4001.  GDVDV and SNM acknowledge support from ANR Grant No. ANR-23-CE30-0020-01 EDIPS. MK and SS thank the hospitality of Laboratoire de Physique Th\'eorique et Mod\`eles Statistiques (LPTMS), Universit\'e Paris-Saclay and Laboratoire de Physique Th\'eoriqueet Hautes Energies (LPTHE), Sorbonne Universit\'e, Paris, France.

\appendix

\section{Calculation of the kernel using Christoffel–Darboux formula}
\label{a:kernel}
In this appendix, we provide details of the calculation of the scaling kernel $K_N(x,y)$, both near the edge and the origin, starting from the Eqs.~\eqref{eq:K+s}--\eqref{eq:KN2} with $M=N/2$ (for simplicity, we take $N$ even; this is immaterial in the large-$N$ limit).

\subsection{Kernel near the edge}
\label{a:edge}

As mentioned in \eref{eq:KNAiry}, 
near the right edge, we expect the following scaling form for the kernel
\begin{equation}
    K_N(x,y)= \frac{1}{W_N} \, K_\text{edge} \left(\frac{x-x_\text{edge}}{W_N}, \frac{y-x_\text{edge}}{W_N}\right).
    \label{eqa:KNAiry}
\end{equation}
The scaling function can be found by evaluating
\begin{equation}
    K_\text{edge}(u,v)=\lim_{N\to\infty} \, W_N\, K_N(x_\text{edge}+W_N\, u, x_\text{edge}+W_N \,v)\,,
    \label{eqa:Kuv}
\end{equation}
where $K_N(x,y)$ is given in \eref{eq:kernel-MB2}, together with Eqs.~\eqref{eq:K+s}--\eqref{eq:KN2}. As discussed in Eqs.~\eqref{eq:z1z20}--\eqref{eq:z1z2}, 
\begin{eqnarray}
    K_\text{edge}(u,v)= \lim_{M\to\infty}\, W_{2M}\, \left[\bar{K}_M^+(z_1,z_2) + \bar{K}_M^-(z_1,z_2)\right],
    \label{eqa:Kuv2}
\end{eqnarray}
where 
\begin{equation}
    W_N = \frac{z_\text{edge}}{\alpha+2}\, N^{-\frac{2\alpha+1}{3(\alpha+2)}}, \quad\text{with}\quad z_{\text{edge}} = \left(\frac{\alpha+2}{\mu}\right)^{\frac{1}{\alpha+2}},
    \label{eqa:WN}
\end{equation}
and
\begin{equation}
    z_1 = 4M + 2 (2M)^{1/3}\,u\, ,\quad\text{and}\quad
     z_2 = 4M + 2 (2 M)^{1/3}\,v\, .
      \label{eqa:z1z2}
\end{equation}

To obtain \eref{eqa:Kuv} with the scaling  \eref{eqa:z1z2}, we use the Plancherel-Rotach asymptotic~\cite{szeg1939orthogonal}, 
\begin{align}
    \text{For ~} z&=4 m + 2 (\beta+1)
 - 2 (2m/3)^{1/3} t\notag\\
 e^{-z/2}\,L_m^{(\beta)}(z) &= (-1)^m 2^{-\beta}\, (2m)^{-1/3}\, \text{Ai}(-3^{-1/3} t) + O (1/m)\, .
 \label{eqa:PR1}
 \end{align}
This can be recast as
\begin{align}
    e^{-z/2}\, L_m^{\beta}(z) = & (-1)^m 2^{-\beta}\, (2m)^{-1/3}\, \text{Ai}\left(\frac{z-4m - 2 (\beta+1)}{2 (2 m)^{1/3}}\right) \notag\\ &+ O(1/m) .
    \label{eqa:PR2}
\end{align}
Setting $z=z_1$ and $z_2$ respectively, and using  \eref{eqa:z1z2}, we get,
\begin{widetext}
\begin{equation}
\begin{split}
    e^{-z_1/2}\, L_M^{\beta}(z_1) & = (-1)^M \left[2^{-\beta -1/3} M^{-1/3}\, \text{Ai}(u)-2^{-\beta -2/3} (\beta +1) \, M^{-2/3}\text{Ai}'(u)\right] + O(1/M)\\
    e^{-z_1/2}\, L_{M-1}^{\beta}(z_1) & =  (-1)^{M+1} \left[2^{-\beta -1/3} M^{-1/3}\, \text{Ai}(u)-2^{-\beta -{2}/{3}} (\beta -1) \, M^{-2/3}\text{Ai}'(u)\right]  + O(1/M) \\
   e^{-z_2/2}\, L_M^{\beta}(z_2) & = (-1)^M \left[2^{-\beta -1/3} M^{-1/3}\, \text{Ai}(v)-2^{-\beta -2/3} (\beta +1) \, M^{-2/3}\text{Ai}'(v)\right] + O(1/M)\\
    e^{-z_2/2}\, L_{M-1}^{\beta}(z_2) & =  (-1)^{M+1} \left[2^{-\beta -1/3} M^{-1/3}\, \text{Ai}(v)-2^{-\beta -{2}/{3}} (\beta -1) \, M^{-2/3}\text{Ai}'(v)\right]  + O(1/M) 
\end{split}
\label{eqa:Ls}
\end{equation}
\end{widetext}
This gives
\begin{align}
    e^{-(z_1+z_2)/2}\,\left[L_{M-1}^{(\beta)} (z_1)\, L_{M}^{(\beta)} (z_2)- L_{M}^{(\beta)} (z_1)\, L_{M-1}^{(\beta)} (z_2) \right] \notag \\=  \frac{2^{-2\beta }  }{M}\, \left[\text{Ai}(u) \text{Ai}'(v)-\text{Ai}'(u) \text{Ai}(v)\right] + o(1/M).
    \label{eqa:LL}
\end{align}
Using \eref{eqa:LL} in Eqs.~\eqref{eq:KN1} and \eqref{eq:KN2}, from \eref{eqa:Kuv2}, we get
\begin{equation}
    K_\text{edge}(u,v) = \frac{\text{Ai}(u) \text{Ai}'(v)-\text{Ai}'(u) \text{Ai}(v)}{u-v}\, .
    \label{eqa:AiryKernel}
\end{equation}

\subsection{Kernel near the origin}
\label{a:kno}

As discussed in \sref{s:korigin}, the typical interparticle separation close to the origin scales as
\begin{equation}
    \ell_0 = \left(\frac{\alpha+2}{\mu}\right)^{\frac{1}{\alpha+2}}\, N^{-\frac{1}{\alpha+2}},
    \label{eqa:l0}
\end{equation}
Therefore, near the origin, we expect the scaling form of the correlation kernel as
\begin{equation}
    K_N(x,y)=\frac{1}{\ell_0}\, K_0\left(\frac{x}{\ell_0}, \frac{y}{\ell_0}\right)\, .
    \label{eqa:KN1}
\end{equation}
The scaling function can be found by evaluating
\begin{equation}
K_0(u,v)=\lim_{N\to\infty} \, \ell_0\, K_N(\ell_0\, u, \ell_0 \,v)\,,
    \label{eqa:Kuv2}
\end{equation}
where $K_N(x,y)$ is given in \eref{eq:kernel-MB2}, together with Eqs.~\eqref{eq:K+s}--\eqref{eq:KN2}. Since the variables $x$ and $y$ in the expression of $K_N(x,y)$ appear through the combination in \eref{eq:z1z20}, substituting $x=\ell_0 \,u$ and $y=\ell_0\, v$, in Eqs.~\eqref{eq:KN1} and \eqref{eq:KN2}, we get 
\begin{equation}
    z_1= \frac{u^{\alpha+2}}{M} \equiv \frac{\tilde{z}_1}{M} \quad\text{and}\quad  z_2= \frac{v^{\alpha+2}}{M} \equiv \frac{\tilde{z}_2}{M}
    \label{eqa:z2z2t}
\end{equation}
where $M=N/2$ and for simplicity, we take $N$ even. This is immaterial in the large-$N$ limit.

To evaluate Eqs.~\eqref{eq:KN1} and \eqref{eq:KN2} in the large $M$ limit, we recast the
Mehler–Heine formula~\cite{abramowitz1965handbook, szeg1939orthogonal}
\begin{equation}
    \lim_{m\to\infty} m^{-\beta} L_m^{(\beta)}(z/m)= z^{-\beta/2}\, J_\beta (2\sqrt{z}),
      \label{eqa:M-H}
\end{equation}
as
\begin{align}
    L_m^{(\beta)} (\tilde{z}/M) &= L_m^{(\beta)} \left(\frac{(m/M)\,\tilde{z}}{m}\right) \notag \\ &\simeq m^\beta\, [(m/M) \, \tilde{z}]^{-\beta/2}\, J_\beta\left(2 \sqrt{(m/M)\, \tilde{z}}\right),
    \label{eqa:M-H2}
\end{align}
where $J_\beta(x)$ is the Bessel function of the first kind. Now, setting $m=M$ or $M-1$, respectively, we get
\begin{align}
    \lim_{M\to\infty}& M^{-2\beta+1} \left[L_{M-1}^{(\beta)} (z_1)\, L_{M}^{(\beta)} (z_2)- L_{M}^{(\beta)} (z_1)\, L_{M-1}^{(\beta)} (z_2) \right]\notag\\
    &= 
      (\tilde{z}_1 \tilde{z}_2)^{-\beta/2} \Bigl[\sqrt{\tilde{z}_1} J_{\beta }\left(2 \sqrt{\tilde{z}_2}\right) J_{\beta +1}\left(2 \sqrt{\tilde{z}_1}\right) \notag\\
      & \qquad \qquad \quad -\sqrt{\tilde{z}_2} J_{\beta }\left(2 \sqrt{\tilde{z}_1}\right) J_{\beta +1}\left(2 \sqrt{\tilde{z}_2}\right)\Bigr],
      \label{eqa:LLa}
\end{align}
where $z_1,z_2$ and $\tilde{z}_1,\tilde{z}_2$ are related by \eref{eqa:z2z2t}. Using \eref{eqa:LL} in Eqs.~\eqref{eq:KN1} and \eqref{eq:KN2}, from 
\eref{eqa:Kuv2}, we get
\begin{equation}
    K_0(u,v) = K_0^+ (u,v)+ K_0^-(u,v),
\end{equation}
where $K_0^\pm(u,v)$ are given by
\begin{align}
   & K_0^\pm(u,v) = \frac{\alpha+2}{2}\, \frac{(\tilde{z}_1\, \tilde{z}_2)^{\gamma/2}}{\tilde{z}_1-\tilde{z}_2}\, \notag\\
    &\times \left[\sqrt{\tilde{z}_1} J_{\beta }\left(2 \sqrt{\tilde{z}_2}\right) J_{\beta +1}\left(2 \sqrt{\tilde{z}_1}\right)-\sqrt{\tilde{z}_2} J_{\beta }\left(2 \sqrt{\tilde{z}_1}\right) J_{\beta +1}\left(2 \sqrt{\tilde{z}_2}\right)\right],
\end{align}
where $\beta=\mp\gamma$ for $K_0^\pm(u,v)$ respectively, and we recall from \eref{eqa:z2z2t} that $\tilde{z}_1=u^{\alpha+2}$ and $\tilde{z}_2=v^{\alpha+2}$.

\bibliography{references}

\end{document}